\documentclass[
 reprint,
 superscriptaddress,
 amsmath,amssymb,
 prd,
floatfix,
]{revtex4-2}

\usepackage{graphicx}% Include figure files
\usepackage[caption=false]{subfig}
\usepackage{footnote}
\usepackage{dcolumn}% Align table columns on decimal point
\usepackage{bm}% bold math
\usepackage{array}% table things
\usepackage[hidelinks]{hyperref}% add hypertext capabilities
\usepackage{svg}%svg file support 
\usepackage{xr}

\usepackage[
separate-uncertainty = true,
multi-part-units=single
]{siunitx}
\usepackage{xcolor}
\usepackage[version=4]{mhchem}
\usepackage[normalem]{ulem}
\usepackage{tablefootnote}
\usepackage[capitalise]{cleveref}
\usepackage{multirow}

\usepackage{xfrac}
\usepackage{lineno}

\usepackage[version=4]{mhchem}

\newcolumntype{C}[1]{>{\centering\let\newline\\\arraybackslash\hspace{0pt}}m{#1}}

\newcommand{\plus}{\raisebox{.2\height}{\scalebox{.8}{+}}}

\DeclareSIUnit\keVnr{keV_{nr}}
\DeclareSIUnit\keVee{keV_{ee}}

\newcommand\UChRn{${}^{222}$Rn}
\newcommand\UCh{${}^{238}$U}
\newcommand\ThCh{${}^{232}$Th}
\newcommand\UChPoE{${}^{218}$Po}
\newcommand\UChPoL{${}^{214}$Po}

\newcommand\ThChRn{${}^{220}$Rn}
\newcommand\ThChPoE{${}^{216}$Po}
\newcommand\ThChPoL{${}^{212}$Po}

\newcommand\UChPb{${}^{214}$Pb}

\newcommand\RnPo{${}^{222}$Rn--${}^{218}$Po~}
\newcommand\PoPb{${}^{218}$Po--${}^{214}$Pb~}
\newcommand\PoBiPo{${}^{218}$Po--${}^{214}$BiPo~}
\newcommand\PbBi{${}^{214}$Pb--${}^{214}$Bi~}

\newcommand\ThChBiPo{${}^{212}$Bi-${}^{212}$Po}
\newcommand\RnPonospace{${}^{222}$Rn--${}^{218}$Po}
\newcommand\PoPbnospace{${}^{218}$Po--${}^{214}$Pb}

\newcommand\Mercury{\textsc{Mercury}}

\begin{document}

\title{Flow-dependent tagging of  \UChPb~decays\\ in the LZ dark matter detector}
% 1 
\author{J.~Aalbers}
\affiliation{SLAC National Accelerator Laboratory, Menlo Park, CA 94025-7015, USA}
\affiliation{Kavli Institute for Particle Astrophysics and Cosmology, Stanford University, Stanford, CA  94305-4085 USA}

% 2 
\author{D.S.~Akerib}
\affiliation{SLAC National Accelerator Laboratory, Menlo Park, CA 94025-7015, USA}
\affiliation{Kavli Institute for Particle Astrophysics and Cosmology, Stanford University, Stanford, CA  94305-4085 USA}

% 3 
\author{A.K.~Al Musalhi}
\affiliation{University College London (UCL), Department of Physics and Astronomy, London WC1E 6BT, UK}

% 4 
\author{F.~Alder}
\affiliation{University College London (UCL), Department of Physics and Astronomy, London WC1E 6BT, UK}

% 5 
\author{C.S.~Amarasinghe}
\affiliation{University of California, Santa Barbara, Department of Physics, Santa Barbara, CA 93106-9530, USA}

% 6 
\author{A.~Ames}
\affiliation{SLAC National Accelerator Laboratory, Menlo Park, CA 94025-7015, USA}
\affiliation{Kavli Institute for Particle Astrophysics and Cosmology, Stanford University, Stanford, CA  94305-4085 USA}

% 7 
\author{T.J.~Anderson}
\affiliation{SLAC National Accelerator Laboratory, Menlo Park, CA 94025-7015, USA}
\affiliation{Kavli Institute for Particle Astrophysics and Cosmology, Stanford University, Stanford, CA  94305-4085 USA}

% 8 
\author{N.~Angelides}
\email{n.angelides@imperial.ac.uk \\ (now at University of Z{\"u}rich, nicolas.angelides@uzh.ch)}
\affiliation{Imperial College London, Physics Department, Blackett Laboratory, London SW7 2AZ, UK}

% 9 
\author{H.M.~Ara\'{u}jo}
\affiliation{Imperial College London, Physics Department, Blackett Laboratory, London SW7 2AZ, UK}

% 10 
\author{J.E.~Armstrong}
\affiliation{University of Maryland, Department of Physics, College Park, MD 20742-4111, USA}

% 11 
% ...

\collaboration{The LZ Collaboration}

\date{\today}
\begin{abstract}
The LUX-ZEPLIN (LZ) experiment is searching for dark matter interactions in a liquid xenon time projection chamber (LXe-TPC). This article demonstrates how control of the flow state in the LXe-TPC enables the identification of pairs of sequential alpha-decays, which are used to map fluid flow and ion drift in the liquid target. The resulting transport model is used to tag \UChPb~ beta-decays, a leading background to dark matter signals in LZ. Temporally evolving volume selections, at a cost of 9.0\% of exposure, target the decay of each \UChPb~ atom up to 81 minutes after production, resulting in (63~$\pm$~6$_{(\mathrm{stat})}$~$\pm$~7$_{(\mathrm{sys})}$)\% identification of \UChPb~decays to ground state. 
We also demonstrate how flow-based tagging techniques enable a novel calibration side band that is concurrent with science data. 
Finally we report updated estimates of radon-chain charge branching fractions in liquid xenon, finding branching to $^{218}$Po$^+$ at $0.49 \pm 0.01$, $^{214}$Pb$^+$ at $0.48 \pm 0.12$, and $^{214}$Bi$^+$ at $0.74 \pm 0.05$, with a mean charged ion lifetime in the LZ TPC of $49 \pm 4$ min.
\end{abstract}

%Use showkeys class option if keyword
                              %display desired
\maketitle

\section{Introduction}
\label{sec:Intro}

Liquid xenon (LXe) time projection chambers have become widespread in rare-event-search particle physics experiments including neutrinoless double-beta decay and direct detection of dark matter \cite{Lebedenko:2008gb, Akerib:2017vbi, PhysRevC.89.015502, Aprile:2015uzo, Si:2022lyh, LZ-TDR}. The LUX-ZEPLIN (LZ) experiment at the Sanford Underground Research Facility (SURF) is a 10 tonne~(t) LXe detector, making it the largest dark matter detector in the world currently in operation. The LXe-TPC contains a 7\,t active mass of LXe in a polytetrafluoroethylene (PTFE) cylinder with inner dimensions of $1.45$\,m in height and diameter with a thin layer of xenon vapor at the top of the detector. The cylindrical TPC is installed inside a titanium cryostat.

Two arrays of 3 inch VUV-sensitive Hamamatsu R11410-22 photomultiplier tubes (PMTs) collect light produced inside the detector; the bottom array is immersed in the cold liquid, and the top array is located in the gas above the liquid surface. The space between the TPC and the cryostat wall, laterally, and in the dome region below the TPC, contains $\sim2$\,t of LXe and is also instrumented with PMTs (Hamamatsu R8520 and R8778),  serving as an active prompt veto referred to as the ``Skin'' detector.

Four electrode grids generate the vertical electric fields needed to drift and subsequently emit ionization electrons into the gas phase \cite{Linehan_2022}. The drift field is established in the ``active region'' between a cathode at the bottom of the detector and a gate grid 5~mm below the liquid surface. Field uniformity is achieved with a set of field-shaping rings embedded within the UV-reflective PTFE cylinder, forming a field-cage enclosing the LXe-TPC. The anode grid located 8~mm above the liquid surface establishes an ``extraction region'' above the gate that ensures cross-phase emission of electrons past the liquid surface and subsequent generation of electroluminescence light in the gas phase.  Additional details on the construction of the LZ LXe-TPC and other detector elements are provided in~\cite{LZ-detector}.

\begin{figure}[t!]
\centering
\includegraphics[width=\columnwidth]{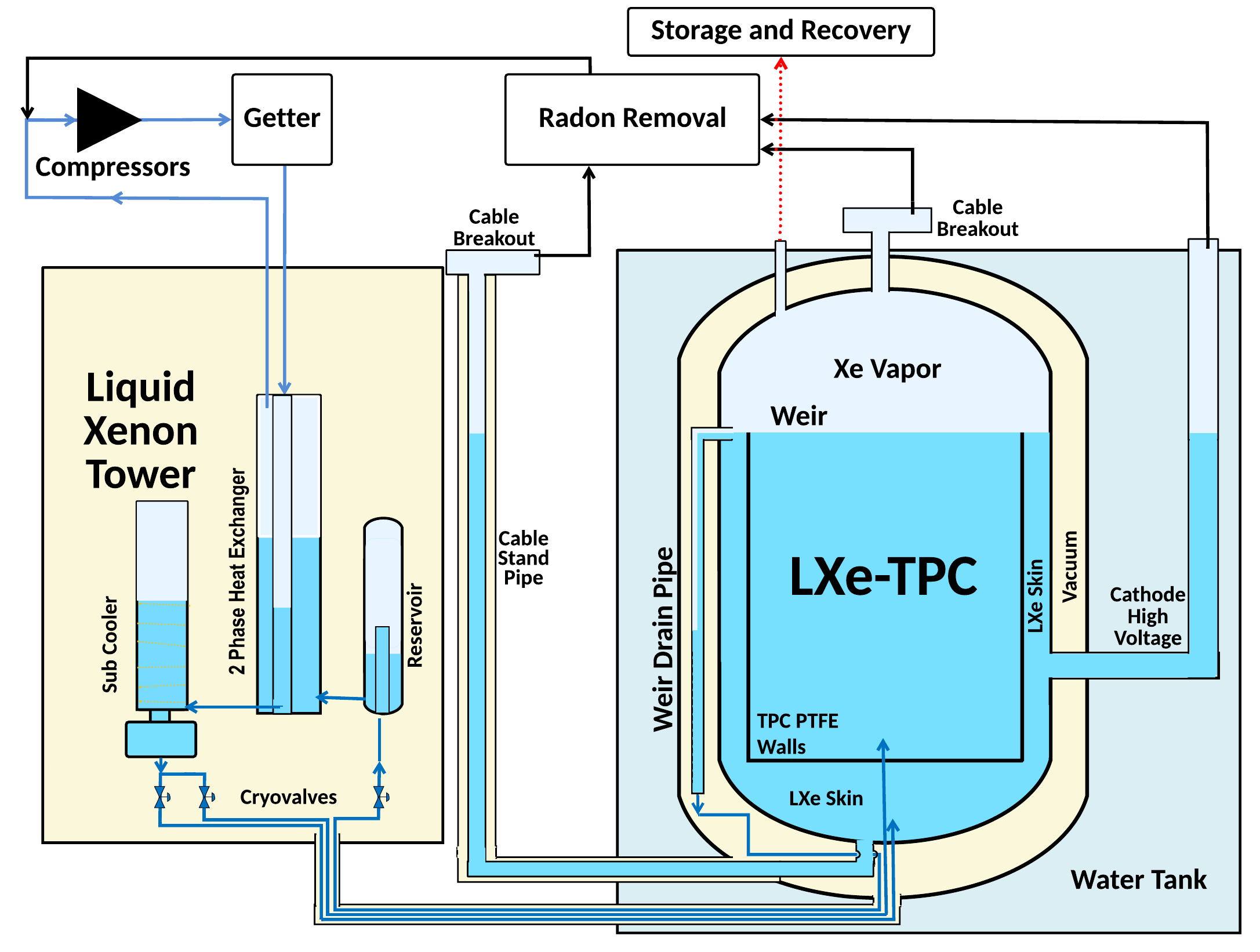}
\caption{A simplified schematic of the LZ xenon circulation system, re-printed with permission from \cite{LZ-detector}.  The subcooler and cryovalves in the Xenon Tower allow experimenters to adjust the temperature and partition the flow of the returning xenon, controlling the flow state in the LXe-TPC as described in the text.
\label{fig:flowdiagram}
}
\end{figure}

The xenon is continuously circulated and purified during operation. The liquid is extracted at the top of the detector, vaporised, purified with a heated zirconium-based purifier,  recondensed, and returned as a sub-cooled liquid at the bottom of the detector (see Figure~\ref{fig:flowdiagram}). Spill-over weirs at the top of the cryostat ensure the stability of the liquid level within the LXe-TPC. Freshly purified LXe enters the cryostat via two delivery lines: one terminating in the dome region of the Skin beneath the TPC and the other running directly to the LXe-TPC through seven inlets arranged in a centered-hexagonal configuration in the bottom PMT array. Separate heaters and cryovalves on the LXe-TPC and Skin inlet lines allow for independent control of the incoming xenon temperature and flow rate for the two spaces, which are separated by the semi-porous field-cage and array structure. 
Additional thermal control is provided by thermosyphons and heaters attached to the upper cryostat wall. 
These controls have enabled exploration of a range of TPC temperature gradients and liquid flow states, from turbulent mixing to coherent circulation. 

Particle interactions within the active volume are observed through prompt scintillation (S1) and delayed electroluminescence (S2) light. The time delay between each S1 and S2 pulse provides the vertical position of an interaction, whereas the S2 hit pattern on the top PMT array is used for horizontal position reconstruction. Nuclear recoil (NR) interactions, such as those expected from the interactions of Weakly Interacting Massive Particles (WIMPs), and electron recoil (ER) interactions, which are background-like in the context of a WIMP search, produce distinct, partly overlapping bands in the S2 vs S1 parameter space \cite{LZ-firstResults}. 

For low-energy rare event searches in LZ ($\lesssim\mathcal{O}(10)$~keV deposited energy in the LXe-TPC), the leading source of background is the $\beta^-$ decay of $^{214}$Pb following \UChRn~emanation from \UCh~series contaminants \cite{LZ-backgrounds}. Trace contamination present in the detector and subsystem materials, particularly in the titanium walls of the vessel containing the LXe-TPC, is responsible for continually producing radon isotopes which emanate and mix with the liquid target. These free radon atoms undergo a sequence of alpha- and beta-decays, with the most prominent of these being \cite{Rn222_ENDSF, Po218_ENDSF, Pb214_ENDSF}:
\begin{align*}
    \ce{^222_86Rn (3.8d) &-> ^218_84Po (Q_\alpha= 5.6MeV) }\,, \\
    \ce{^218_84Po (3.1m) &-> ^214_82Pb (Q_\alpha= 6.1MeV) }\,, \\
    \ce{^214_82Pb (27.1m) &-> ^214_83Bi (Q_\beta= 1.0MeV) }\,, \\
    \ce{^214_83Bi (19.7m) &-> ^214_84Po (Q_\beta =3.3MeV) }\,, \\
    \ce{^214_84Po (164\mu s) &-> ^210_82Pb (Q_\alpha= 7.8MeV) } \, .
\end{align*} 
Each decay in this sequence can produce neutral or singly positively charged progeny \cite{EXO}. Neutral progeny mix in the LXe target through bulk fluid flow and diffusion, while charged progeny also drift due to the vertical electric field.

The beta-decay of \UChPb~to the ground state of $^{214}$Bi (12.7(9)\% branching fraction \cite{Pb214_ENDSF}) is observed as a population of single-scatter ER events, including low-energy events which may partially leak into the NR signal region; this is the dominant low-energy ER background in many rare event searches, including LZ.
For this reason, extensive material screening campaigns~\cite{LZ-screening, lz-rnemanation, XENON1T-screening, PandaX-4T-screening} as well as integrated radon removal systems based on chromatography~\cite{lz-irrs} or distillation~\cite{xenon-irrs, XENON:2025nic, PandaX-RnRemoval} have been developed and applied as the main \UChPb~mitigation efforts to the current generation of LXe-TPC dark matter detectors, achieving \UChRn~levels as low as 0.90~$\mu$Bq/kg~\cite{XENON:2025nic}.  
LZ's mitigation program has resulted in an average activity of $4.78~\pm~0.33~\mu$Bq/kg for \UChRn~in the LXe-TPC (an average of two decays per minute), resulting in a \UChPb~ER background 6$\times$ the irreducible neutrino ER background in the WIMP region of interest (ER energies below $\sim$15~keV)~\cite{LZ-backgrounds}.

 The low \UChRn~activities achievable in LXe-TPCs enable the pairing of consecutive \UChRn~and \UChPoE~decays with spatially and temporally correlated signals. Each pair describes the displacement of charged or neutral progeny (\UChPoE) between production and decay. Observed \RnPo pairs facilitate the study of ion mobility, LXe flow states and their stability, and operational controls over the spatial distribution of radon-chain backgrounds in these experiments.

The observed stability of the liquid flow in LZ during its first science run \cite{LZ-firstResults} has enabled the development of a technique to further suppress the \UChPb~background through flow-based identification, or ``tagging,'' of \UChPb~decays, as described in this paper. Selection volumes defining regions around the predicted paths of \UChPb~neutral atoms and ions are initiated at each progenitor \UChPoE~decay site. Each such volume evolves in size, shape and location, as informed by a liquid flow map and ion mobility. 
Remarkably, the liquid flow is slow and stable enough that, combined with the low radon concentration, the selection volumes result in only a small loss of exposure while achieving selection efficiency for \UChPb~that is more than an order of magnitude greater than previously achieved in the field \cite{XENONtagging_2024}.  This tag has already been employed in LZ's recent dark matter search result~\cite{LZ-WS2024} and is described in detail for the first time in this work.  Note that all data considered in this work were taken at the 193\,V/cm drift field used in LZ's first science run~\cite{LZ-firstResults}, but the analysis techniques developed here apply equally well to data taken at the 97\,V/cm drift field used in later runs (including the data appearing in LZ's 2024 WIMP search~\cite{LZ-WS2024}) with only minor changes to the details of alpha-decay selections and the ion drift model presented in later sections.

The $^{214}$Bi and $^{214}$Po decays that occur after the $^{214}$Pb decay are also used throughout this work.  The $^{214}$Bi half-life is of the same scale as that of $^{214}$Pb, while the $^{214}$Po half-life (164 $\mu$s) is short compared to LZ's millisecond-scale drift times \cite{LZ-firstResults}, resulting in $^{214}$Po decays that usually overlap with the preceding $^{214}$Bi event window. This $\beta^-$-$\alpha$-decay-coincidence provides a unique event topology, referred to in this work as a $^{214}$BiPo event.  Comprehensive and pure $^{214}$BiPo event selections are possible, and a \PoBiPo tag constructed in the same fashion as the \PoPb tag serves to both tune and validate the tagging model. 

Section~\ref{sec:Flow} of this article demonstrates the impact of detector flow state on $^{222}$Rn-related backgrounds even in the absence of flow tagging. Section~\ref{sec:Alphas} demonstrates a more comprehensive alpha-decay selection and position reconstruction scheme, useful in maximizing the impact of flow tagging. Section~\ref{sec:PairFinding} shows how the identified decays of \UChRn~and \UChPoE~are matched into pairs, forming vectors describing ion mobility and flow. The development of the \UChPb~tag is presented in Section~\ref{sec:Tagging} where the path and shape of selection volumes targeting neutral and charged \UChPb~are tuned and the resulting tagging efficiency is measured. Finally, the impact to physics searches is discussed in Section~\ref{sec:Impact}.

\section{Flow State and the Radon-Chain}
\label{sec:Flow}

LZ is able to explore how adjustments to parameters of the circulation system impact the degree of LXe mixing and the overall distribution of \UChRn~and its progeny in the LXe-TPC. The circulation settings considered in this work are constrained to maintain fixed LXe-TPC pressure, liquid surface temperature, and xenon circulation rate.  Given these constraints, two significant free parameters remain: the temperature of the xenon returning to the detector and the distribution of LXe delivery between the LXe-TPC and LXe Skin.

\begin{figure*}[t!]
\centering
    \includegraphics[width=\columnwidth*2, trim = 0 0 0 0, clip=true]{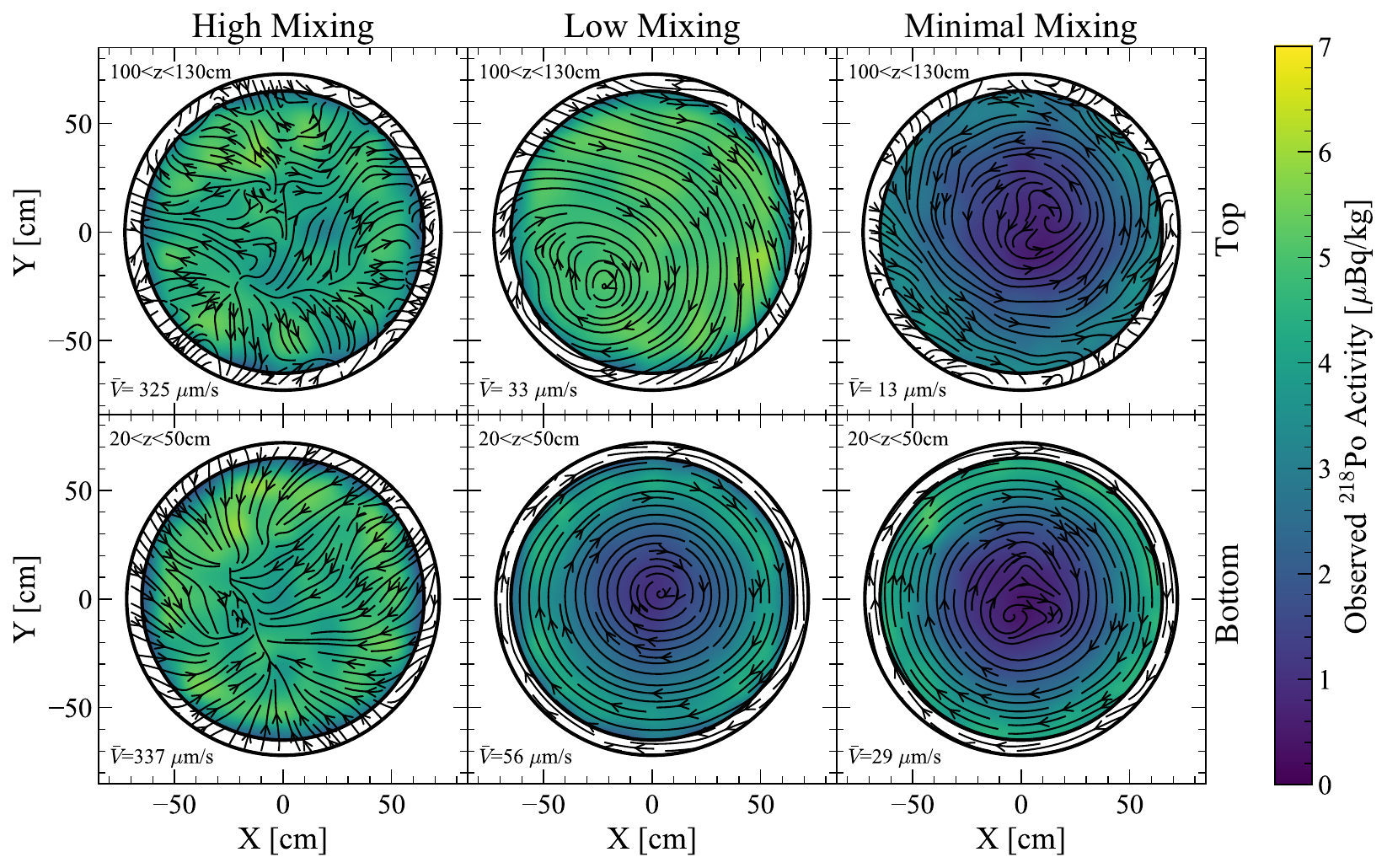}
    \caption{Comparison of three flow states with different degrees of mixing in the LXe-TPC. From left to right the columns correspond to: High Mixing, Low Mixing, and Minimal Mixing, whereas the two rows represent a higher and lower slice of the LXe-TPC (outer black circle). In the coordinate system shown, the cathode at the bottom of the active region is located at Z=0~cm and the gate grid at the top is located at Z=145.6~cm. Arrows demonstrate the observed horizontal LXe flow as determined by pairs of \RnPo~decays with $\Delta T >$~5~s. The activity of \UChPoE, shown within the 5.5~t fiducial volume (inner circle), is indicated by color, with a fixed color scale across all flow states and slices. An average speed is also indicated in each panel, calculated from \RnPo~pairs with at least one decay position observed within that slice. For the Low and Minimal Mixing states, only ``neutral'' pairs are used as input to this figure. In the High Mixing state separate ``neutral'' and ``charged'' pair populations are not evident and no attempt is made to exclude ``charged'' pairs from the selection.  The analysis presented in Sections~\ref{sec:PairFinding} and \ref{sec:Tagging} is based on data taken in the Low Mixing state.
    \label{fig:mixing_states}
    }
\end{figure*}

Three distinct mixing states (referred to throughout this paper as High Mixing, Low Mixing, and Minimal Mixing) and their impact on \UChRn~progeny are shown in Figure~\ref{fig:mixing_states}, constructed using the alpha-decay selections and \RnPo pair-finding techniques described in Sections~\ref{sec:Alphas} and ~\ref{sec:PairFinding}, respectively. 
The High Mixing state is achieved by increasing the temperature of the LXe entering the bottom of the detector to $\sim$2~K above the liquid surface temperature, with increased cooling power applied at the top of the cryostat to maintain fixed LXe-TPC pressure. 
In this highly convective state LXe mixing has proved sufficient to distribute both \UChRn~atoms and internal calibration sources uniformly within the LXe-TPC, so the High Mixing state has become the default state for calibration source injection~\cite{LZ-cal}.

In direct contrast, in the Low Mixing state the LXe entering the bottom of the detector is $\sim$3~K below the liquid surface temperature, establishing a stable temperature gradient and suppressing vertical convection currents. This circulation state was used in LZ's first science run~\cite{LZ-firstResults} and is now the default state for taking science data in LZ.
In this state the LXe at the bottom of the detector settles into circular flow, reducing the likelihood that \UChRn~emanated from the cryostat walls reaches the LXe fiducial volume. 
This circular flow pattern breaks down in the upper half of the LXe-TPC, introducing \UChRn~into the fiducial volume, which allows $^{218}$Po$^+$ ions to spread throughout the vertical range of LZ. This reduces the  effectiveness of flow-based \UChRn~segregation as a background control measure, resulting in only a 10$\%$ reduction in fiducial $^{218}$Po activity relative to the highly mixed state \cite{LZ-backgrounds}.

Finally, in the Minimal Mixing state the temperature of incoming xenon is decreased to $\sim$4~K below the surface temperature, and the distribution of incoming LXe is rebalanced to increase flow to the LXe-TPC (with a corresponding decrease in flow to the Skin). This configuration was explored briefly following LZ's first science run, in an attempt to encourage an overall outwards radial flow, further reducing the ingress of radon-contaminated LXe to the inner regions of the detector. The Minimal Mixing state exhibits coherent circular flow lacking any radial currents along the entire height of the detector, with an observed reversal in the direction of circular flow, from clockwise to anticlockwise as viewed from above, around half way up the LXe volume.
This state achieves nearly 20\% average reduction in the activity of \UChPoE~within the 5.5~t fiducial volume compared to the Low Mixing state, as shown in Table~\ref{tab:rn_mixing_activities}. Assuming otherwise identical conditions, this would correspond to a reduction of $\sim$15\% in \UChPb~activity, derived through a simple charged-ion-drift Monte-Carlo time-step algorithm as done in Ref.~\cite{LZ-backgrounds}.

\begin{table}[t]
    \centering
    \setlength{\extrarowheight}{2pt}  
    \begin{tabular}{|c|c|c|c|}
    \hline
        \UChPoE~activity & Entire TPC & Fiducial & Inner Fiducial \\\relax
        [$\mu$Bq/kg] & (7~tonne) & (5.5~tonne) & (3~tonne) \\
        \hline
        High Mixing  & $5.0 \pm 0.3$ & $5.1 \pm 0.2$ & $5.1 \pm 0.2$  \\
        Low Mixing  & $4.8 \pm 0.3$ &  $4.6 \pm 0.1$ & $3.9 \pm 0.2$   \\
        Minimal Mixing   & $4.2 \pm 0.3$ & $3.9 \pm 0.2$ & $3.1 \pm 0.1$  \\
        \hline
    \end{tabular}
    \caption{Activities of \UChPoE~measured within the LZ LXe-TPC in different sub-volumes.  Each volume has diameter equal to its height and is centered in the middle of the active region.}
    \label{tab:rn_mixing_activities}
\end{table}

The common detector pressure of the different mixing states minimizes the operational impact of transitions between states. Each circulation change to a state of lower mixing requires  one to two weeks to stabilize, while the High Mixing state can be achieved within a day. The Low and Minimal Mixing states show stability over time with deviations in flow appearing only if LXe circulation is disturbed, and the Low Mixing state has been found to be effectively reproducible, with no observed change in $^{214}$Pb tagging performance once circulation settings are restored.

The flow models for the Low Mixing and Minimal Mixing states are conditioned on \RnPo pair densities of 1~pair per 27~cm${}^3$ (one to two months exposure), selecting from the neutral \RnPo pair population based on vertical velocity as described in Section~\ref{sec:PairFinding}.
The High Mixing flow state, on the other hand, demonstrates a limitation in \RnPo pair finding and flow modeling when LXe flow speeds approach or exceed the charged ion drift speed of $\mathcal{O}(0.1)$~mm/s.  In this state, no separation in vertical velocity is observed between neutral and charged \RnPo pair populations.  It is unknown whether this difficulty disentangling xenon flow from charged ion drift is due solely to spatial variation in the vertical flow velocity, in which case a higher-density sample of flow vectors than is currently available could in principle allow flow tagging in the High Mixing state, or to temporal and possibly turbulent variation as well, in which case there may be a fundamental limitation to flow tagging in highly-mixed states.

The exposure required to constrain a flow model makes it desirable to select a single flow state for science data taking.  The Low Mixing state has been selected in this regard for multiple reasons, including the stability and reproducibility of the Low Mixing state demonstrated in LZ's first science run~\cite{LZ-firstResults} and the modest time required to reestablish the state following calibration source injections and other flow interruptions.  All data used in the remainder of this work, and the  majority of the livetime in all LZ science runs to date~\cite{LZ-firstResults, LZ-WS2024,LZ:2025igz}, has been taken in the Low Mixing state.

\section{Reconstruction and Identification of Alpha Decays}
\label{sec:Alphas}

Radon atoms produced in detector materials emanate out of surfaces, mixing with the liquid xenon before undergoing a sequence of decays. However, some of the decays of interest may take place after an atom attaches to a surface of the detector, resulting in charge loss and subsequent degradation of energy and position reconstruction.
In this section we explain how the positions and energies of alpha decays are reconstructed within the liquid bulk, inside or outside the active volume and on detector surfaces, in order to identify individual decays.

The two relevant series from which alpha decays are observed in the LZ LXe-TPC are the \UCh~and \ThCh~series, populating the \UChRn~and \ThChRn~sub-chains. This work focuses on the former,  which includes \UChPb. The \ThCh~decay series, observed at much lower activity ($\lesssim$~10\%) than the \UChRn~chain, is suppressed due to the much shorter half-life of \ThChRn~(56~s) \cite{Nucl-Data-rn220, LZ-backgrounds}.

Alpha decays with energy in the 5--8~MeV range produce short ($\lesssim$100~$\mu$m) tracks with high ionization density in LXe \cite{xe-tracks-marrodan}. Despite the applied drift field of 193\,V/cm in the active region, recombination in the track results in a much larger number of primary scintillation photons (S1$>$10,000~photons detected) than ionization electrons (S2 $>900$~electrons detected)~\cite{Bradley:2014,Rischbieter:2022,Aprile:2006kx}. 

When both S1 and S2 signals corresponding to an alpha decay are observed, the decay position is reconstructed with the S2 time delay and hit pattern mapping onto vertical and horizontal positions, with resolutions of 0.7~mm and 4~mm, respectively \cite{LZ-firstResults}. However, for events where a decay occurs outside the active region (no S2), where charge is lost or degraded due to saturation or field effects (compromised S2), or where either spatially or temporally separated energy depositions are present (multiple S2s), information about position is extracted purely through the S1.

While less precise than positions derived using S2, the S1-based position reconstruction is necessary both to extend the transport model developed in Section~\ref{sec:PairFinding} to the edges of the active region, and to reconstruct the $^{214}$BiPo events used to build tagging windows in Section~\ref{sec:Tagging}.  These applications are tolerant of the lower resolution and potential for misconstruction inherent in S1-based positions.  Where accuracy and precision are needed, in particular for $^{222}$Rn and $^{218}$Po decays in the 5.5~t fiducial volume, over 99.5\% of decays include a reliable S2 signal.

The primary method for reconstructing position from S1 invokes the same \Mercury~\cite{LUX-Position, merc_og} algorithm used to reconstruct horizontal position from the S2.
This algorithm uses the S1 PMT hit-map from both arrays in conjunction with position dependent PMT light response functions 
to infer the $XY$ and $Z$ coordinates of the decay.
In cases where \Mercury~does not converge, a simpler alternative is used which reduces the full S1 hit-map on both arrays to three parameters: the centroid $X$ and $Y$ coordinates and the ratio of light between the two arrays (top-bottom asymmetry, or TBA). 
S1-TBA-based $Z$ positions rely on a simple 1-D interpolation from TBA to $Z$, calibrated on alpha decay events with a good S2 signal. S1-centroid-based $XY$ positions use a Delaunay triangulation \cite{delaunay1934sphere}, as is implemented in the CERN ROOT library \footnote{ROOT::Math::Delaunay2D Class Reference \url{https://root.cern.ch/doc/master/classROOT_1_1Math_1_1Delaunay2D.html}}, to construct a continuous, piecewise linear map from the centroid $X$ and $Y$ to physical $XY$ coordinates, again calibrated on a set of alpha S1 signals for which accurate positions are known from the presence of a good S2 signal. 

The precisions of the S1-based position reconstruction algorithms are assessed within the active region of the detector where S2-derived position information is accurate. The difference between S1-derived and S2-derived positions forms a Gaussian distribution, the mean and standard deviation of which indicate each S1 algorithm's residual systematic bias and uncertainty. The uncertainty on the \Mercury-based S1-derived position was found to be 1 cm in all axes, whereas the uncertainty on the simpler parametric S1-derived position was found to be 3~cm in the horizontal plane and 1~cm in the vertical direction. Systematic bias effects were found to be of the order of 1~mm for all S1-derived positions.  
The resolutions quoted here are averages over the detector volume, with some degradation in vertical resolution at the top and bottom of the TPC.

For every candidate event, the S1-based reconstructed \textit{Z}-position is used in determining and verifying the matching S2. If a ``good" S2 is found within five standard deviations of the expected drift time then position is reconstructed using that S2. If no S2 is found, or if the S2 found appears diminished or distorted, the S1-based reconstructed position is used. Once the decay position is determined, S1s are corrected for spatially-dependent light collection efficiency effects.
For events reconstructed to regions beyond the existing S1 correction tables, the closest known correction factor is used in combination with a small additional TBA-dependent correction.
This corrected S1 is used later as a linear energy scale, fitting the most prominent peaks to known alpha-decay energies. 

\begin{figure}[t]
	\centering
	\includegraphics[width=\columnwidth]{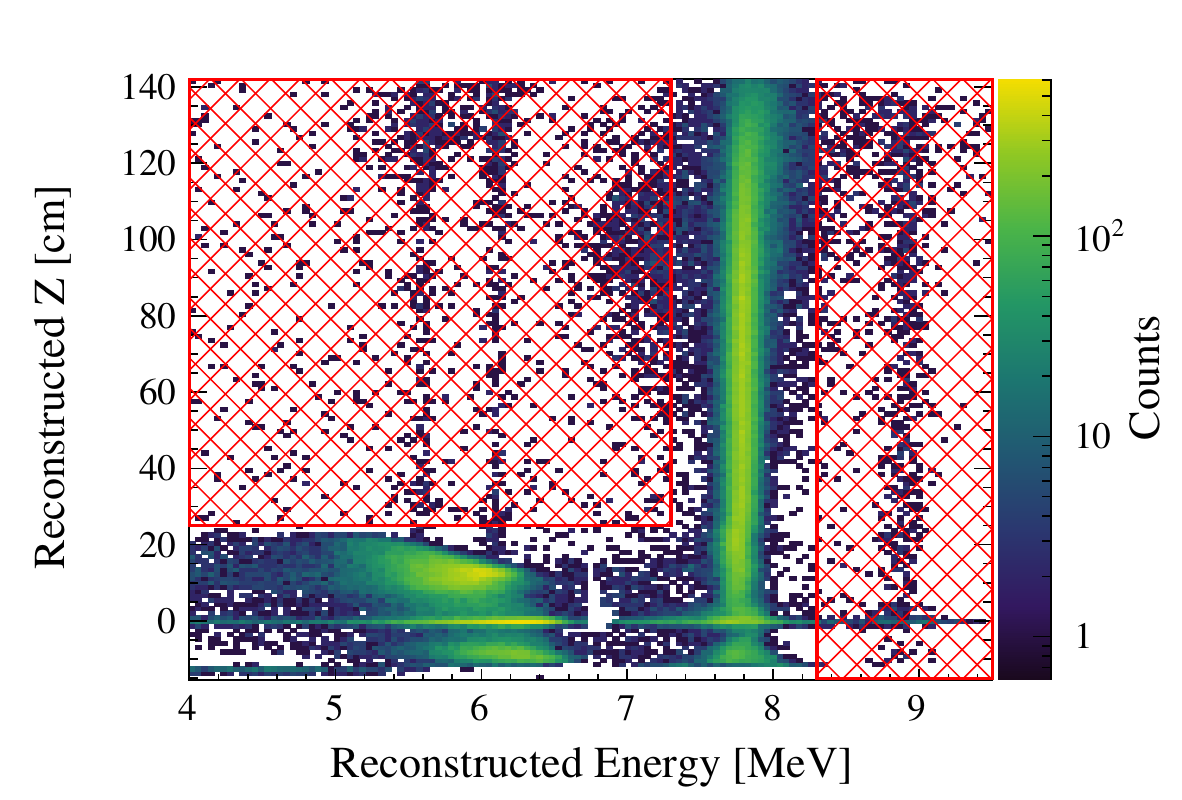}
	\caption{Reconstruction of events with two S1 pulses, with the larger S1 following the smaller within 650~$\mu$s, designed to target the beta-alpha delayed coincidence of $^{214}$BiPo events. The energy scale is linear in the position-corrected size of the second S1.  The red-grid regions contain accidental coincidences between background events and $^{222}$Rn or $^{218}$Po, seen around 5.6~MeV and 6.1~MeV respectively, as well as \ThChBiPo~coincidences seen near 9~MeV, and are excluded from the $^{214}$BiPo selection. The populations at or near 0~cm at around 6~MeV are believed to correspond to the cathode grid $^{214}$BiPo population. 
    The reduction in reconstructed energy for some $^{214}$BiPo events at \textit{Z} of 60--120~cm is due to a data acquisition saturation effect, impacting events where gamma emission from the $^{214}$Bi decay produces multiple S2 pulses prior to the S1 from the $^{214}$Po decay. 
    } \label{fig:bipos}
\end{figure}

The position reconstruction performance can be explored through an analysis of $^{214}$BiPo events, which can be independently identified by their unique multi-S1 event topology. The $^{214}$BiPo event selection takes events containing at least one pair of S1 pulses with the larger S1 following the smaller within 650~$\mu$s (four times the half-life of the \UChPoL~decay). Events passing this rough selection are shown in Figure~\ref{fig:bipos}. In addition to the bulk population observed at the decay energy of 7.8~MeV, this selection enables the identification of a population of \UChPoL~atoms on the cathode grid (lower boundary of the active region). The complex light collection efficiency around the cathode wires results in the mis-reconstruction of both energy and vertical position, reconstructing most events to $\sim$6~MeV with apparent positions ranging from 25~cm above to 15~cm below the cathode. With this information, data selections to identify cathode events from any radon-chain decay are developed. 
Further detail on the signals associated with $^{222}$Rn-chain decays on and in the cathode grid wires is given in~\cite{LZ:2026hpq}.

\begin{figure}[t]
	\centering
	\includegraphics[width=0.5\textwidth]{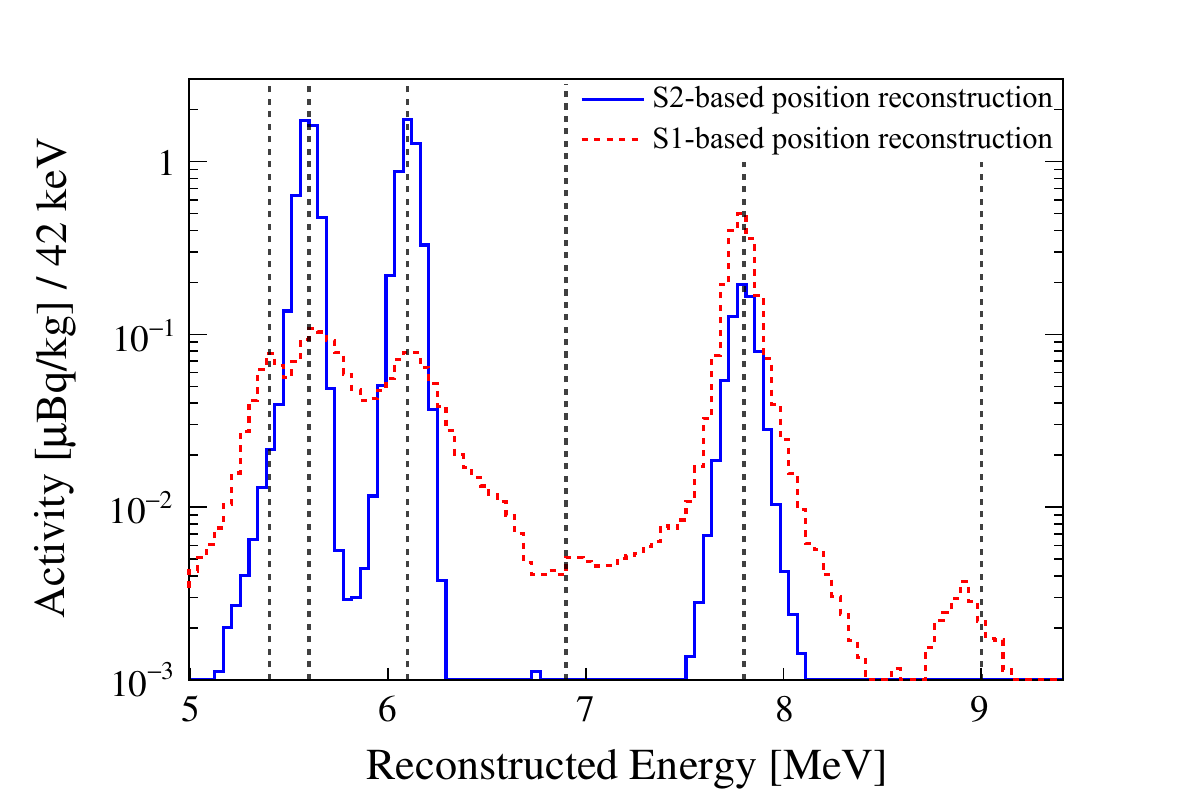}
	\caption{Reconstructed energy spectra for alpha decays from the two dominant radon chains in the LZ LXe-TPC, excluding cathode populations, in the Low Mixing state.  The energy scale is linear in position-corrected S1, and vertical lines mark the literature values for the alpha-decay $Q$-values, from left to right: $^{210}$Po, \UChRn, \UChPoE, \ThChPoE, \UChPoL, and \ThChPoL. Populations reconstructed with S1-based (red dashed line) and S2-based (solid blue line) position information are shown separately. The first ($^{210}$Po) peak appears exclusively in the S1-based reconstruction because those decays occur on the TPC walls.  Short half-life decays such as $^{214}$Po (5th peak) and $^{212}$Po (6th peak) also rely heavily on S1-based reconstruction due to S2 pile-up. The $^{212}$Po peak appears slightly below the expected energy due to signal saturation affecting any decay over 8.5~MeV.
    Corrected energy resolutions (standard deviation divided by mean based on a Gaussian fit to each peak) for $^{214}$Po are 0.89\% and 1.15\% for events with S2-based positions and S1-based positions, respectively, improving to 0.87\% and 0.99\% for events in the fiducial volume.} 
	\label{fig:corr_space}
\end{figure}

Once cathode alpha populations are identified, the remaining alpha decays from the bulk exhibit good resolution in position-corrected S1, which maps linearly to alpha-decay energy as shown in Figure~\ref{fig:corr_space}. In this space, Gaussian fits are performed for prominent peaks separately for populations reconstructed with S1-based and S2-based position information. Alpha decays are identified using these fits, with selections extending out to five standard deviations or to the midpoint of overlapping peaks.

The energy and position reconstruction described above assists the identification of all relevant radon chain alpha decays throughout the active volume. This allows the transport model construction described in Section~\ref{sec:PairFinding} to proceed with no ``fiducial'' cut  applied except where explicitly noted. The evaluation of the resulting tag in Section~\ref{sec:Tagging} does apply a fiducial cut, restricting analysis to regions where S2-based reconstruction is possible. Within the fiducial volume, the energy-based alpha decay identification described above is assumed to be of sufficient efficiency and purity that misidentification of alpha decays is negligible compared to the other known inefficiencies discussed.

\section{$^{214}$P\lowercase{b} Transport Model Construction}
\label{sec:PairFinding}

The transport of radon-chain progeny is subject to both liquid xenon fluid flow and charged ion displacement in the TPC electric field. Successful modeling of each effect is key to predicting later-chain progeny decay locations. 
This section constructs models of both effects through the identification of \RnPo pairs.

After defining some general considerations for pair searches in Section~\ref{subsec:GeneralPairing},
Section~\ref{subsec:InitialPairs} builds a preliminary selection of \RnPo pairs separated by less than 10 minutes.  Section~\ref{subsec:InitialPairValidation} validates this selection, and Section~\ref{subsec:Paraview} uses the initial set of pairs to construct a flow model, which in turn allows the identification of \RnPo pairs with increased purity and efficiency and at larger time separation.  This process is iterated once more to obtain the final flow model used in later sections.
The \RnPo pairs identified here extend to a time-coincidence window of only 15 minutes, but the transport models based on these pairs enable predictions at longer timescales.  Model performance out to 81 minutes is evaluated in Section~\ref{sec:Tagging} by investigating \PoBiPo and \PoPb pairing efficiencies.

\subsection{Preliminary Considerations for Decay Pairing}
\label{subsec:GeneralPairing}

Two populations emerge in each of the decay pairings discussed in this section and in Section~\ref{sec:Tagging}, corresponding to the neutral and $+1$ charge states of the progeny atom/ion. Other charge states are expected to be too short-lived to meaningfully impact progeny spatial populations \cite{Emily_Thesis}.
 Neutral \RnPo pairs identified in this section form the basis of the xenon flow model, while charged pairs inform the ion drift model.

Any pair identification algorithm is expected to produce both true progenitor-progeny decay pairs and mismatched or ``unrelated'' pairs.  Characterizing the amount and distribution of unrelated pair contamination is critical to understanding an algorithm's performance and to extracting flow and mobility uncertainties in the resulting transport model.  Throughout the following analysis, pure samples of unrelated pairs are produced by performing unphysical pair searches, where the progeny event population is artificially translated forward in time by $\mathcal{O}(10)$ half-lives relative to the progenitor population, with the pairing algorithm otherwise unchanged. This avoids all truly related pairs while preserving the underlying progenitor and progeny spatial distributions.  Unrelated pairs can then be statistically subtracted from the nominal pair search, with the remaining ``pair excess'' reflecting the distribution of correctly paired decays.  In this section, unrelated \RnPo pairs are characterized by 40-minute-shifted searches, normalized by the number of $^{222}$Rn progenitors in each search.  Longer time shifts are used to characterize unrelated \PoPb and \PoBiPo pairs in Section~\ref{sec:Tagging}.  Independent samples of the unrelated population given by different time shifts can also be averaged to reduce the statistical uncertainty on the pair excess when that uncertainty is dominated by the unrelated pair subtraction, as in Section~\ref{subsubsec:voxeltag}.

\subsection{Bounds on $^{218}$Po Displacements for Initial Pairing}
\label{subsec:InitialPairs}
\begin{figure}[!t] 
    \centering
    \includegraphics[width=\columnwidth]{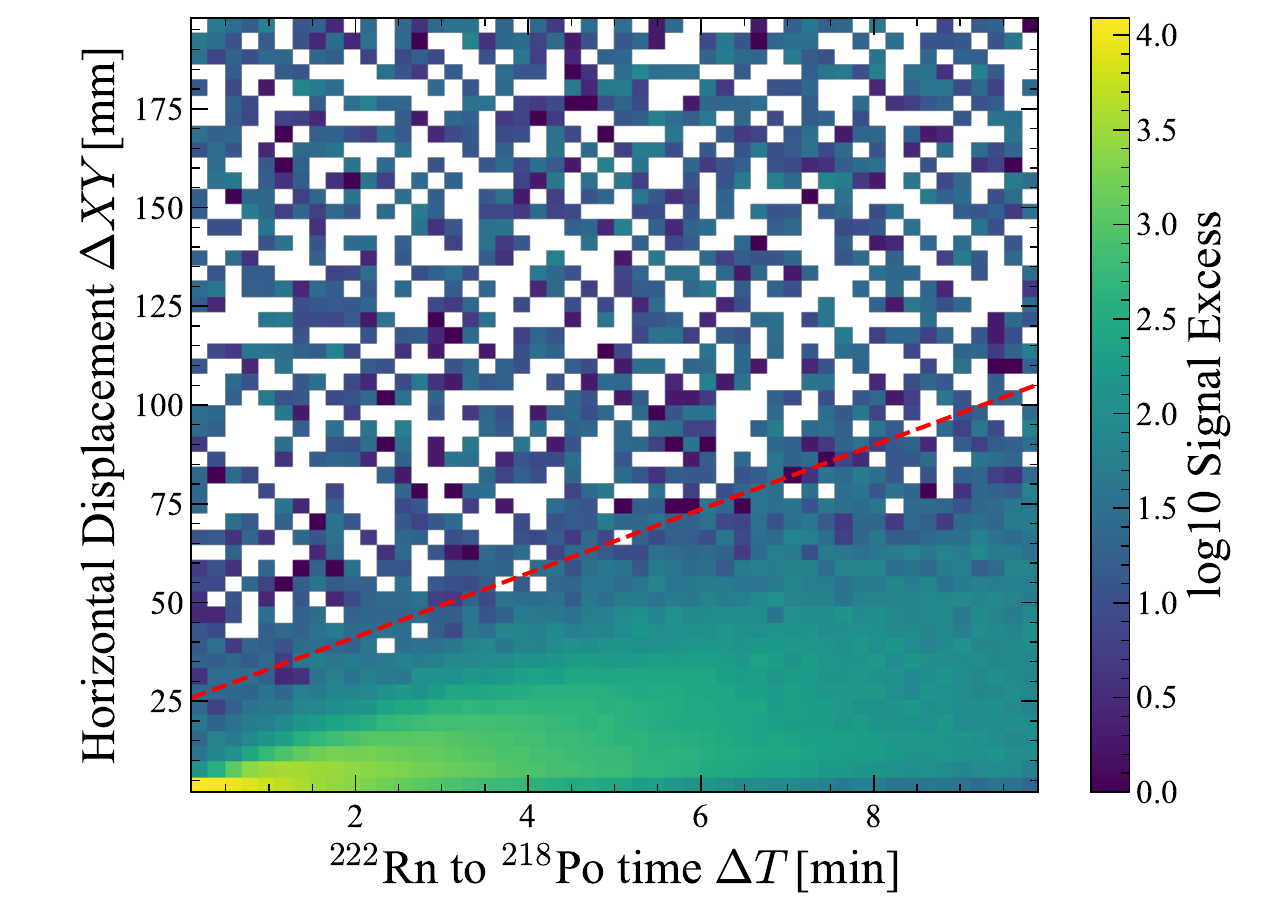} 
    \includegraphics[width=\columnwidth]{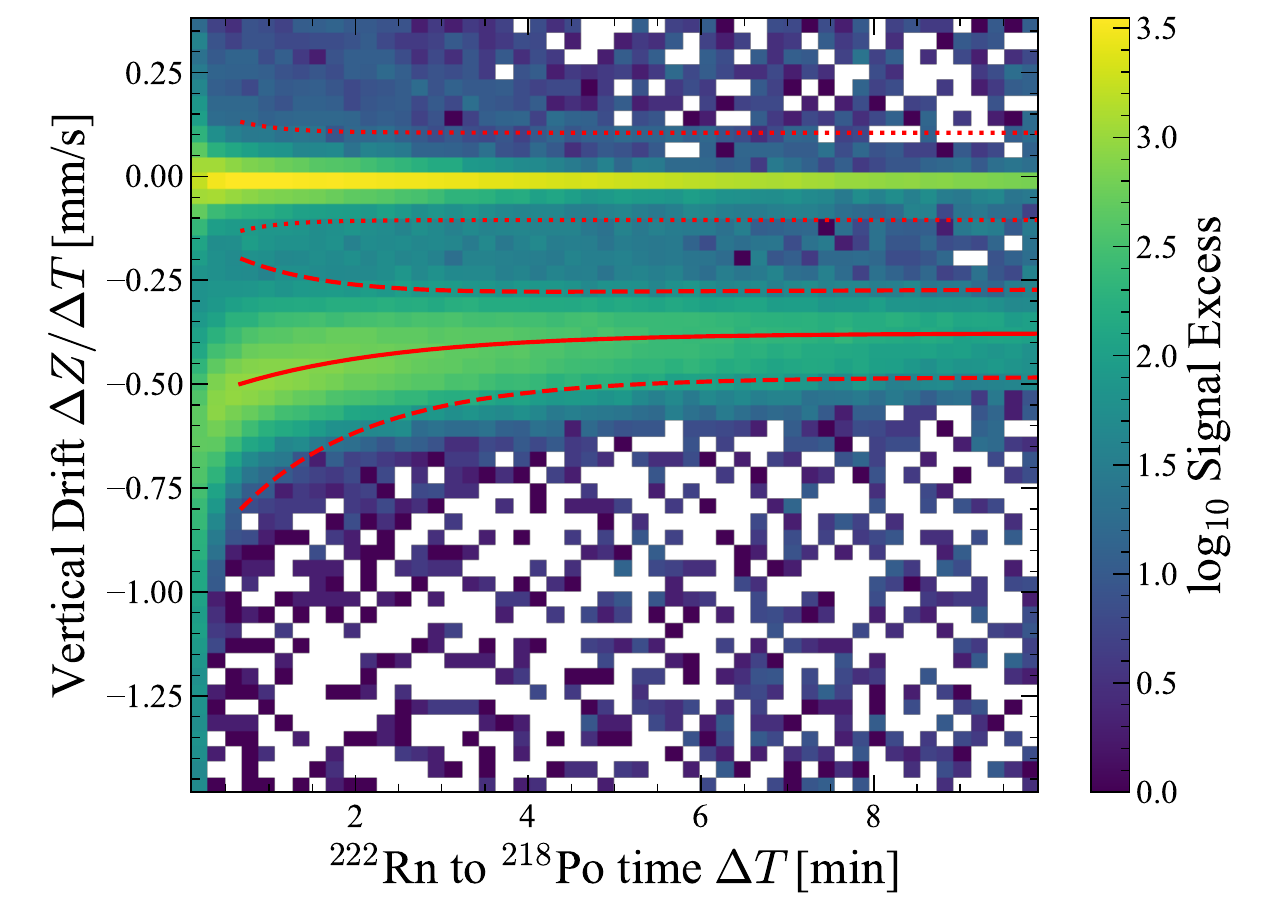}
    \caption{Above: Horizontal displacements of all possible \RnPo pairs out to a 10-minute time separation. The dashed red line corresponds to Eq.~(\ref{EQ:RadialSelection}). Correctly paired decays are observed at small horizontal displacements.  Below: Average vertical velocities of \RnPo pairs passing the horizontal displacement selection indicated in the top panel. The two populations observed correspond to neutral and positively charged \UChPoE~atoms/ions, where the ions move downwards due to the electric field in the LXe-TPC. The solid red line shows $\overline{V}_Z^+$ from Eq.~(\ref{EQ:Charge_Initial_Pairing_Cut}), fit to charged band median values, while the dotted and dashed red lines indicate the pair acceptance cuts defined in Eqs.~(\ref{EQ:Neutral_Initial_Pairing_Cut}) and (\ref{EQ:Charge_Initial_Pairing_Cut}) for neutral and charged progeny, respectively. The charged and neutral selection bands in the figure are truncated at 40 seconds for visual clarity, but selection windows are defined down to $\Delta T=0$.  Both panels apply a  subtraction of the unrelated pair background, estimated from a 40-minute-shifted \RnPo search.}
    \label{fig:IntialRnPoSelections}
\end{figure}

If the radon decay rate is sufficiently low and transport is sufficiently slow, a preliminary, high-purity set of \RnPo pairs can be made even without a transport model (\emph{i.e.}, with a null transport model).  Figure~\ref{fig:IntialRnPoSelections} illustrates this for the Low Mixing state, showing distributions based on the spatial ($\Delta X$, $\Delta Y$, $\Delta Z$) and time ($\Delta T$) separations between all possible \RnPo pair combinations with $0 < \Delta T < 10$~minutes.  Figure~\ref{fig:IntialRnPoSelections} also shows the sequential cuts in this space used to select \RnPo pairs for transport model building.  The first selection is on the horizontal plane with $\Delta XY \equiv \sqrt{\Delta X^2 + \Delta Y^2}$, taking
\begin{equation} \label{EQ:RadialSelection}
\Delta XY < a + b \Delta T, 
\end{equation}
with $a=25$~mm and $b = 0.14$~mm/s, for $\Delta T<10$~minutes.  Two selections are then made on $\Delta{Z}/\Delta{T}$, targeting neutral and charged pairs, respectively.  The neutral selection is given by
\begin{equation} \label{EQ:Neutral_Initial_Pairing_Cut}
\left|\Delta Z / \Delta T\right| < c + d \exp(-\Delta T/\tau),
\end{equation}
with $c= 0.10$~mm/s, $d=0.10$~mm/s and $\tau=30$~s, while the charged selection is given by
\begin{equation} \label{EQ:Charge_Initial_Pairing_Cut}
\begin{aligned}
    \left| (\Delta Z / \Delta T) - \overline{V}_Z^\textbf{\plus}\right| &< e + f \exp(-\Delta T/\zeta), \\ 
     \overline{V}_Z^\textbf{\plus} &= V_L + (V_I-V_L) \exp(-\Delta T/\lambda).
\end{aligned}
\end{equation}
Here $ \overline{V}_Z^\textbf{\plus}$ can be interpreted as the drift velocity of charged progeny averaged over progeny lifetime. The corresponding instantaneous drift velocity, used in the ion transport model, is given by
\begin{equation}
\label{EQ:InstantaneousIonDriftVelocity}
    V_Z^+ = V_L + (V_I-V_L)\left(1-\frac{\Delta T}{\lambda}\right)\exp(-\Delta T/\lambda).
\end{equation}
The width of the charged selection is somewhat wider than the neutral selection, with $e=0.11$~mm/s, $f=0.32$~mm/s, and $\zeta=80$~s, and the center of the selection reflects the decrease in ion drift speed with time, with an initial velocity $V_I=-0.55$~mm/s decaying to an asymptotic velocity of $V_L=-0.38$~mm/s with decay time constant $\lambda=115$~s.  Note that while both the neutral and charged selection windows grow in $\Delta Z$ over time, they do so slower than linearly, giving shrinking windows in $\Delta Z / \Delta T$.  This is reflected in the clear separation between the neutral and charged populations at high $\Delta T$ seen in Figure~\ref{fig:IntialRnPoSelections}.

The time evolution of ion drift velocity seen above has been qualitatively reproduced at both higher and lower electric fields:  LZ's 2024 WIMP Search  \cite{LZ-WS2024} measured ion drift velocities of $V_I=-0.47$ mm/s and $V_L=-0.24$mm/s at 97~V/cm,  while EXO-200 reports initial and final ion drift velocities at 380~V/cm of 1.48~mm/s and 0.83~mm/s, respectively \cite{EXO-200chargedions}. Together these studies suggest a non-trivial time evolution of ion mobility with complex field dependence, but are consistent with a common long-time $^{218}$Po$^+$ ion mobility of $22\pm2\,\textrm{mm}^2\,\textrm{s}^{-1}\,\textrm{kV}^{-1}$. 

The widths of the selection windows in $\Delta XY$ and $\Delta Z/\Delta T$ are determined ``by hand'' based on the regions with significant pair excess over the unrelated pair background as seen in Figure~\ref{fig:IntialRnPoSelections}, and are used only to select pairs for the first iteration of model construction below.  The instantaneous ion drift velocity $V_Z^+$ is used in later stages of the analysis, and is found by fitting the lifetime-averaged ion drift velocity $\overline{V}_Z^+$ to the median of the charged-pair band after a cut similar to that given by Eq.~(\ref{EQ:Charge_Initial_Pairing_Cut}).  
Note that predicted ion positions at long times depend primarily on the asymptotic ion drift velocity $V_L$, and are relatively insensitive to the chosen functional form for $V_Z^+$.

\subsection{Validation of Initial \RnPo Pair Construction}
\label{subsec:InitialPairValidation}
\begin{figure}[!t] 
    \centering
    \includegraphics[width=\columnwidth]{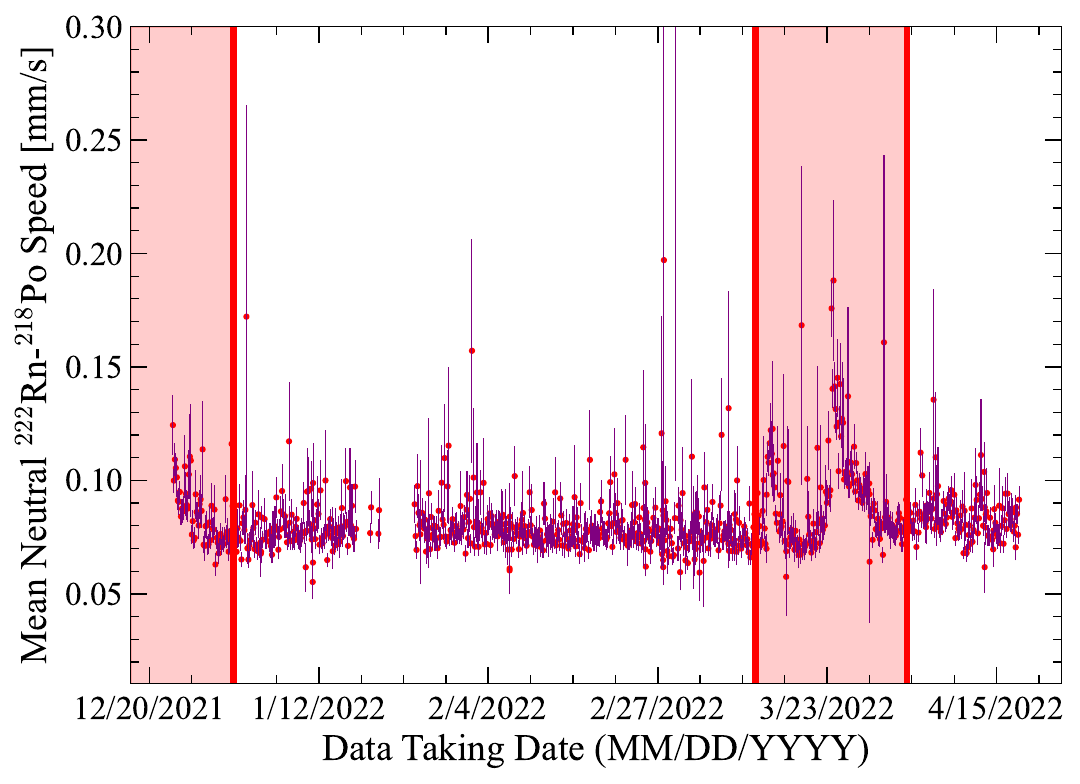}
    \centering
    \caption{Time dependence of the measured average speed of neutral long lived ($ \Delta T>150$ s) \RnPo~decay pairs.  Error bars indicate the statistical uncertainty on the calculated mean, with typical sample sizes of 400 pairs per data point.  No fiducial cut is applied.  Data in the red shaded regions are excluded from this analysis, as the thermodynamic state of the detector was different enough to have a non-negligible impact on flow and pairing efficiency.  Note that the velocities here show the by-pair average, while the velocities indicated in Figure~\ref{fig:mixing_states} are a volume-weighted average.}
    \label{fig:DistinctFlow}
\end{figure}

The xenon flow in the detector can be characterized by the neutral pairs selected using Eqs.~(\ref{EQ:RadialSelection},\ref{EQ:Neutral_Initial_Pairing_Cut}), with an additional cut requiring $\Delta T>150$~s applied to avoid overlap between the charged and neutral populations and to suppress noise from position reconstruction uncertainty.  Figure~\ref{fig:DistinctFlow} shows the average velocity $\left\langle\sqrt{\Delta X^2 + \Delta Y^2 + \Delta Z^2} / \Delta T\right\rangle$ of this population over 5 months of calendar time, identifying two windows with speeds systematically higher than the typical $80\pm20$~$\mu$m/s. The first is due to the transition from commissioning circulation settings to science run settings, and the second is a recovery period following a gas circulation compressor malfunction. These time periods, while still exhibiting slower velocities than in the High Mixing state referenced in Sec.~\ref{sec:Flow}, are excluded from the pairing analysis here and in Sec.~\ref{sec:Tagging}. A similar process was used to identify periods of high flow velocity, high mixing, or general flow state instability in LZ's 2024 WIMP search result~\cite{LZ-WS2024}. 

The purity and efficiency of \RnPo pair selection in the remaining time windows can be determined by statistically subtracting pairing rates in the time-shifted search, indicating that 95.5\% of the pairs in this initial \RnPo selection are true progenitor-progeny pairings.  This corresponds to a pairing rate of 0.755$\pm$0.002~true pairs per $^{222}$Rn decay detected in the LXe-TPC, or 0.820$\pm$0.003~true pairs per $^{222}$Rn decay detected in the 5.5~t fiducial volume.  Most of the observed inefficiency can be accounted for by known effects, including the $\Delta T$$<$10~min requirement on pairs, the LZ data-acquisition system live fraction of 95.9$\pm$0.4\%~\cite{LZDAQ}, and the transport of charged $^{218}$Po$^+$ to the TPC cathode (referred to as ``cathode plate-out'').  Combined, these effects predict a maximum pairing efficiency of 0.830$\pm$0.010 pairs per $^{222}$Rn decay detected in the LZ-TPC and 0.836$\pm$0.010 pairs per $^{222}$Rn decay detected in the 5.5~t fiducial volume, where in both cases the uncertainty is dominated by uncertainty in the cathode plate-out rate.  This indicates that the selection volumes defined above contain 91$\pm$1\% (98$\pm$1\%) of observable progeny out to 10 minutes following $^{222}$Rn decay in the LXe-TPC (fiducial volume).  

Spatial variations in pairing efficiency were also studied and, as suggested by the efficiency numbers above, regions near the walls and electrode grids demonstrate significantly reduced pairing. This is likely driven by a combination of two effects: surface attachment (plateout) preventing alpha-decay detection or interfering with the free vertical displacement of charged ions, and event reconstruction errors affecting decays of both neutral and charged progeny very close to or on a detector surface. 
A similar analysis of the final \RnPo pair selection purity and efficiency is given in Sec.~\ref{subsec:chargemodel}, with a comparison of the observed and expected pair $\Delta T$ distributions shown in Fig.~\ref{fig:ParaViewLifetimeFits}.

\subsection{Construction of the Xenon Flow Model}

\label{subsec:Paraview}

 \begin{figure}[t]
     \centering
     \includegraphics[width=\columnwidth]{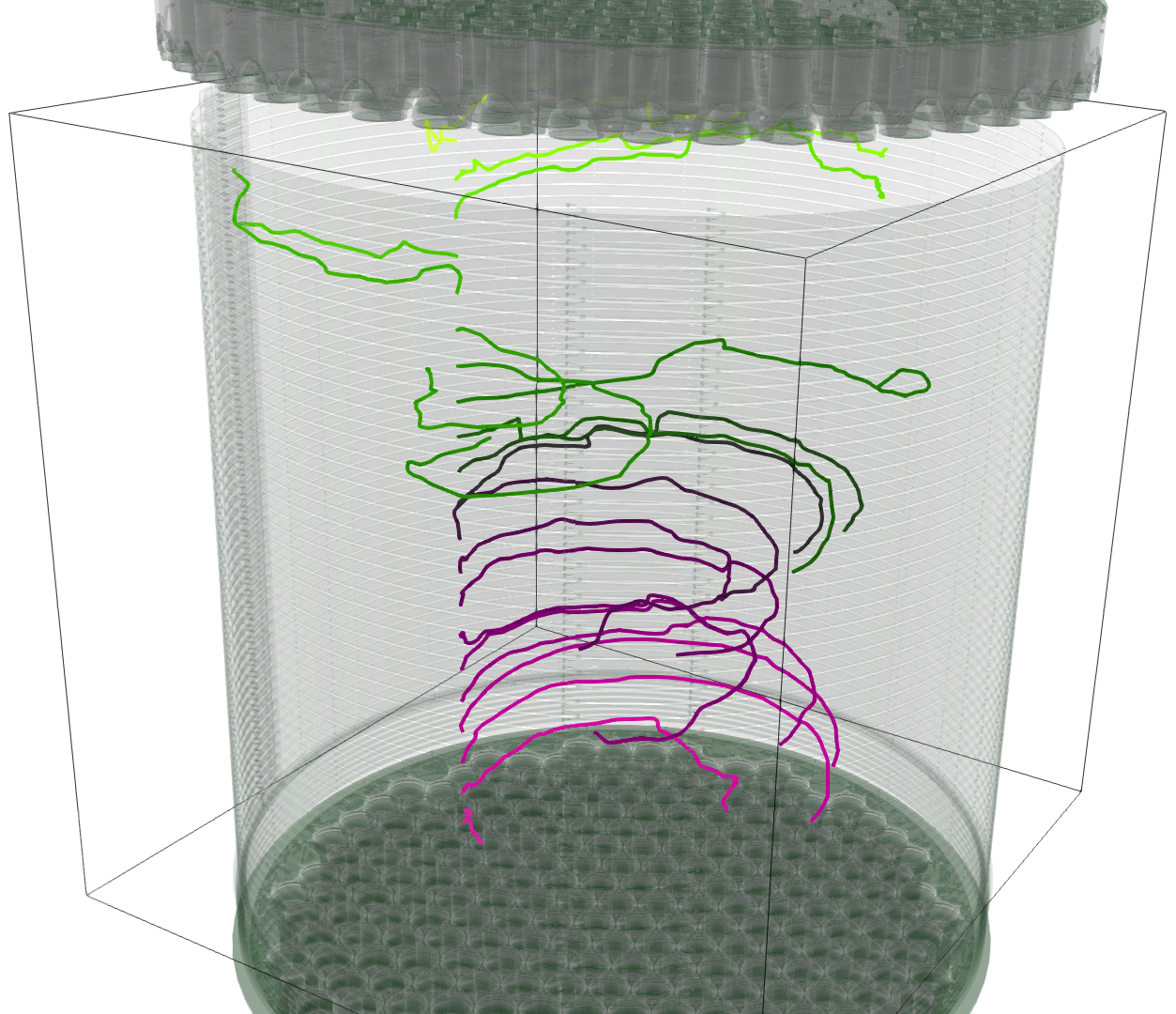}
     \centering
     \caption{Visualization of predicted trajectories (aka ``streamlines'') of neutral atoms in the LXe-TPC, with CAD renderings of the field cage, top PMT array, and bottom PMT array.  The displayed streamlines are projected from initial positions with the same $XY$ ($X=-250$~mm, $Y=-180$~mm; left of the central axis roughly midway to the TPC wall, in the perspective shown) at fixed intervals in $Z$. Each streamline near the bottom is evolved for $\sim$8~hours, whereas streamlines near the top are evolved for up to $\sim$12~hours, both times selected for visual purposes. Streamline color is used only to distinguish the separate streamlines. 
     Well-behaved circular flow is visible at the bottom of the LXe-TPC, while the top shows more complicated flow geometry with radial currents as depicted in the top-center panel of Figure~\ref{fig:mixing_states}.}
     \label{fig:StreamlineVis}
 \end{figure}

The \RnPo pairs selected by the neutral pair window with a time separation $\Delta T>150$~s provide a high-purity sample of xenon flow inside the LXe-TPC.  Each \RnPo pair in this sample is considered to provide a single flow velocity vector, giving the flow velocity at the $^{222}$Rn decay location.  Constructing a usable flow model from this sample requires both an interpolation scheme to estimate flow at arbitrary locations in the LXe-TPC, and an integrator to propagate streamlines in the interpolated flow field.  Both of these functions are supplied by the open source ParaView software package \cite{ParaView}.

Spatial interpolation of the flow field is performed using an isotropic, 10-cm radius Gaussian kernel weighting the 11 nearest sample flow vectors, generating flow velocities on a 9.6-mm grid filling the LXe-TPC.  This structured grid of velocity vectors is then the basis for interpolation at arbitrary points as requested by the streamline integrator, which uses ParaView's Runge-Kutta 4.5 algorithm to project streamlines forward with an adaptive step size of 0.2---5~mm ($<10$~s time steps).  Example streamlines are shown in Figure~\ref{fig:StreamlineVis}.
Projected paths for charged progeny are built by making repeated calls to the ParaView integrator, adjusting the streamline position after each call (i.e. every few seconds in drift time) by the ion drift velocity given in Eq.~(\ref{EQ:InstantaneousIonDriftVelocity}).  Finally, a simple linear interpolation in time is used to evaluate position along either streamline (neutral or charged) at arbitrary time.

\begin{figure}[t] 
    \centering
    \includegraphics[width=\columnwidth]{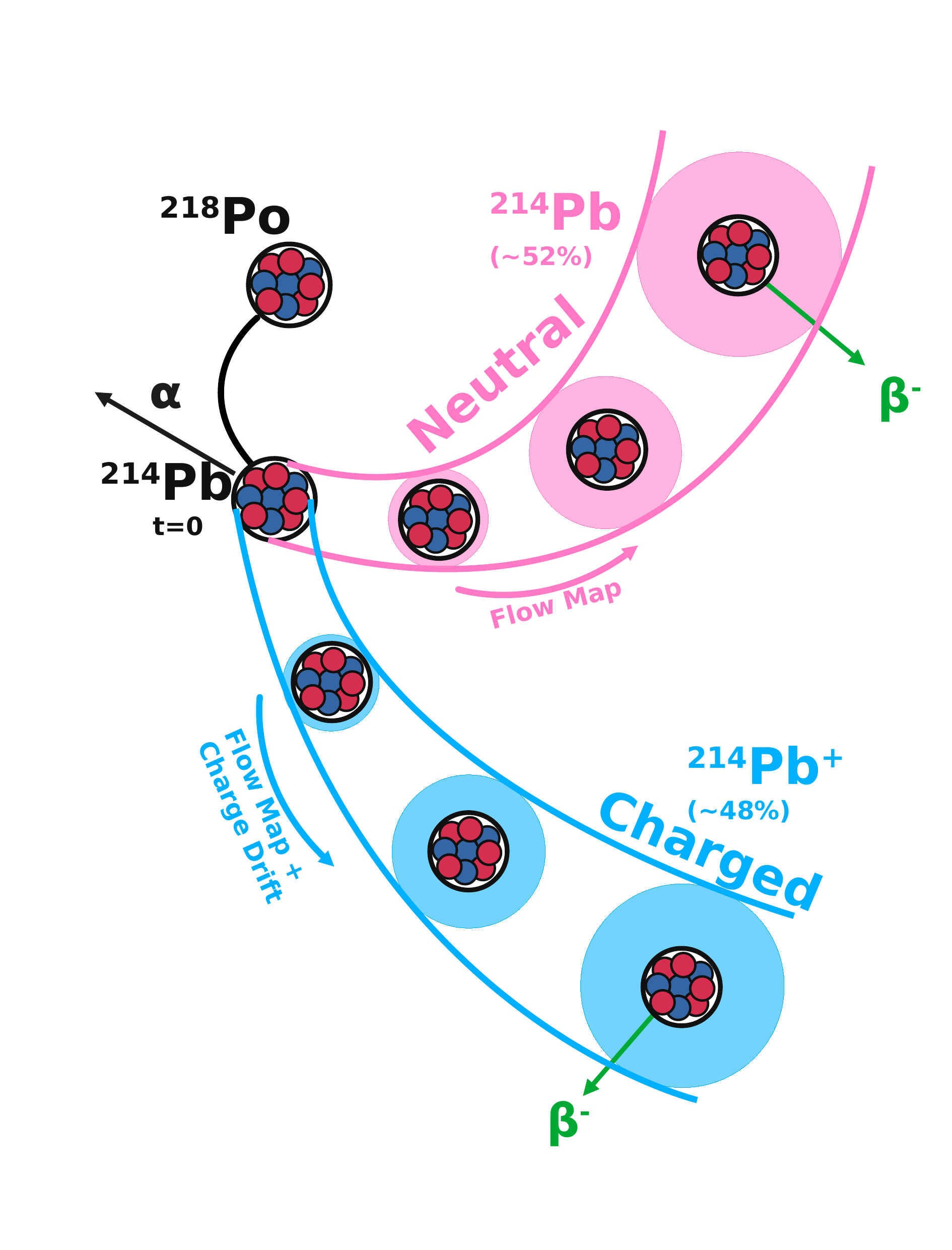}
    \centering
    \caption{Illustration of a \PoPb search extended from the progenitor $^{218}$Po decay along neutral (pink) and charged ion (blue) trajectories. As the best estimate of the \UChPb~position moves along the streamline the search volume for the \UChPb~decay (shaded pink/blue) expands. The rate of expansion is chosen to reflect the flow uncertainty along the path, as defined by Eq.~(\ref{EQ:StreamCut}). }
    \label{fig:StreamlineCartoon}
\end{figure}

Progeny decay positions can now be compared to projected positions on streamlines rather than to progenitor positions (e.g. $\Delta X_\mathrm{s}\equiv X - X_\mathrm{s}$, where $X_\mathrm{s}$ is the projected position on a streamline at the time of the candidate progeny decay).  The preceding \RnPo pair analysis is then repeated, with the progenitor-progeny deviations $(\Delta X, \Delta Y,\Delta Z)$ replaced by two copies of streamline deviations $(\Delta X_\mathrm{s}, \Delta Y_\mathrm{s}, \Delta Z_\mathrm{s})$, one relative to the projected neutral trajectory and one relative to the  charged trajectory as illustrated in Figure~\ref{fig:StreamlineCartoon}.  The resulting distributions allow for tighter selection windows, improving purity from 95.5\% in a 10-minute search to 98.5\% in a 15-minute search, with the extended time window increasing the overall pairing efficiency by 6\% by capturing longer-lived progeny.  This refined \RnPo selection is then used to construct a new flow model, following the identical procedure to the initial flow model.  A further iteration of this process is found not to improve performance.  All references to streamline deviations in the following sections take for a transport model this once-refined \RnPo xenon flow model and the unchanged ion drift model in Eq.~(\ref{EQ:InstantaneousIonDriftVelocity}).

\section{Flow Tag Design, Validation, and Performance}
\label{sec:Tagging}

This section applies the transport model constructed from \RnPo pairs in Section~\ref{sec:PairFinding} to later-chain progeny, looking for both \PoBiPo and \PoPb pairs.  The former, like the \RnPo pairs of the previous section, benefit from the high-efficiency selection of both progenitor (now $^{218}$Po) and progeny ($^{214}$BiPo) events.  Section~\ref{subsec:PoBiPo} takes advantage of this high-efficiency selection by using observed \PoBiPo pair distributions to define pair-selection windows out to 81 minutes (three $^{214}$Pb half-lives), at which point the accidental pairing rate roughly equals the correct pairing rate.  Section~\ref{subsec:PoPb} applies these windows to a \PoPb search, using the $^{214}$Pb background model of \cite{LZ-backgrounds} to determine the resulting $^{214}$Pb tagging efficiency.  Finally, Section~\ref{subsec:chargemodel} uses the charged-daughter branching and ion neutralization measurements described in Appendix~\ref{sec:Supplemental} to model the observed tag efficiencies and $\Delta T$ distributions of the \PoPb and \PoBiPo pair searches.

\subsection{Tuned Tag Windows From \PoBiPo Pairs}\label{subsec:PoBiPo}
The definition of selection windows for \PoBiPo pairs around streamlines extended from $^{218}$Po decays follows a scheme similar to that used for \RnPo pairs in Section~\ref{sec:PairFinding}:  distributions of $\Delta XY_\mathrm{s}$ and $\Delta Z_\mathrm{s}$ versus $\Delta T$ are produced, with the unrelated pair rate statistically subtracted, and selection windows are constructed around the observed excess pair rate.  Unrelated pair rates are larger than for \RnPo pairs due to the longer half-lives involved (see Figure~\ref{fig:ParaViewLifetimeFits} at the end of this section), and are determined using a 24-hour shifted search. Selection windows are defined separately for neutral and charged streamlines.

\begin{figure*}[t]
    \centering
    \subfloat{
    \includegraphics[width=\columnwidth]{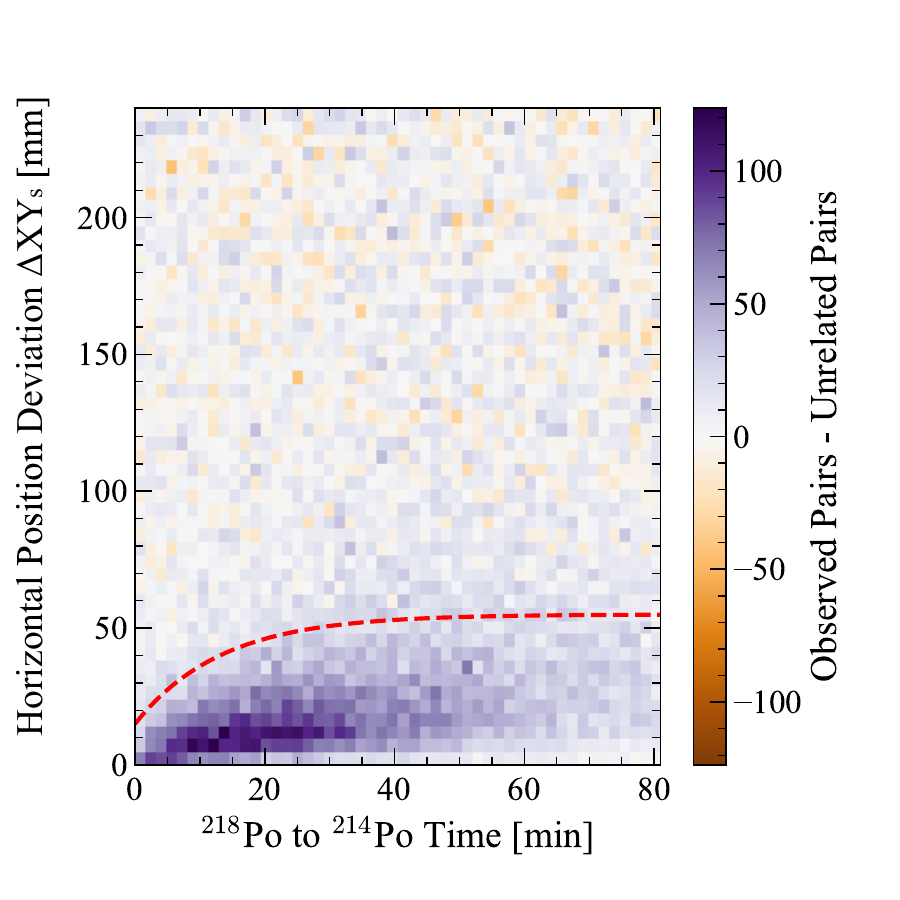}
    \includegraphics[width=\columnwidth]{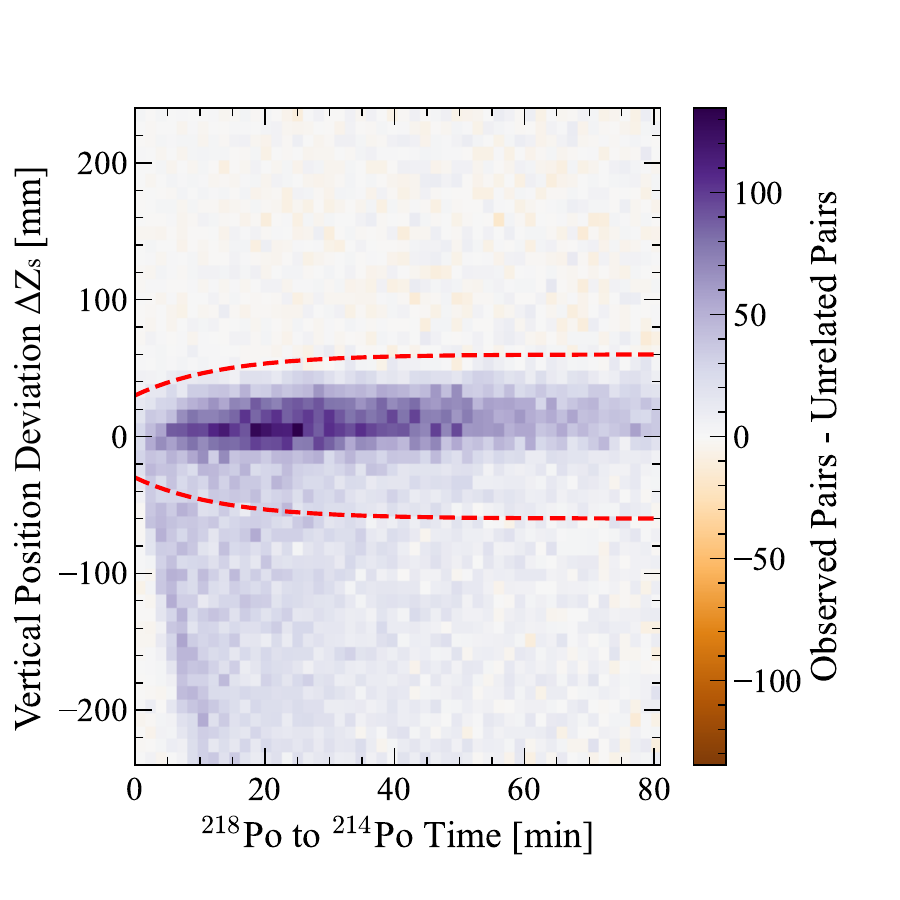}
    }\\
    \begin{tabular}{C{\columnwidth} C{\columnwidth} } 
    A & B
    \end{tabular}
    \caption{\PoBiPo neutral progeny search performance, showing horizontal (panel A) and vertical (panel B) spatial deviations between observed $^{214}$BiPo event locations and the locations projected by neutral streamlines extended from $^{218}$Po decays.  Red dashed lines indicate the selection window defined in Eq.~(\ref{EQ:StreamCut},\ref{EQ:StreamParamsNeutral}), and both panels have applied the selection in the complementary coordinate (\emph{e.g.}, panel A shows horizontal deviations with the $\Delta Z_{s}$ cut applied).  Both panels have statistically subtracted the unrelated pair background, determined from a 24-hour-shifted search.  Clear excesses over unrelated pair rates are seen within the selection windows.
    Charged and semi-charged pairs found within the neutral streamline's horizontal acceptance window are also visible as an excess at negative vertical deviation in panel B. 
    Statistical uncertainties in the background-subtracted rate can be estimated from the color-spread in the regions where the average pair excess is zero, with typical per-bin statistical uncertainties of 11 and 6 pairs in panels A and B respectively.
    }
    \label{fig:Paraview_Neutral_Tuning}
\end{figure*}

\begin{figure*}[p]
    \centering
    \subfloat{
    \includegraphics[width=\columnwidth]{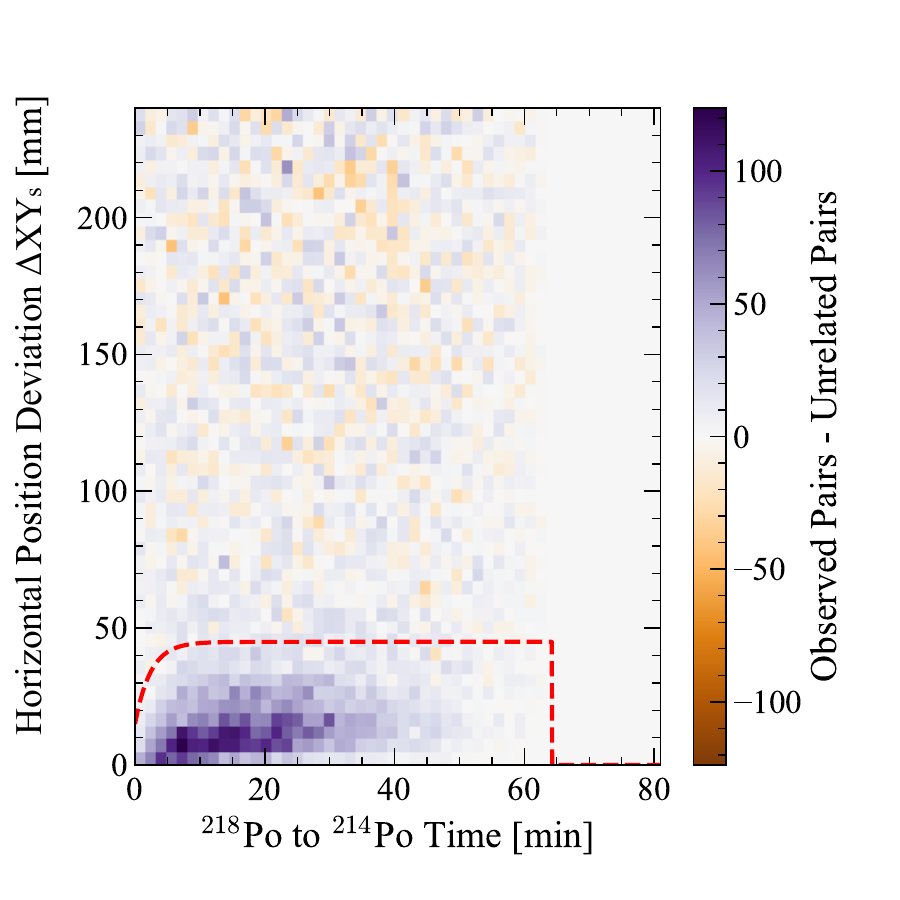}
    \includegraphics[width=\columnwidth]{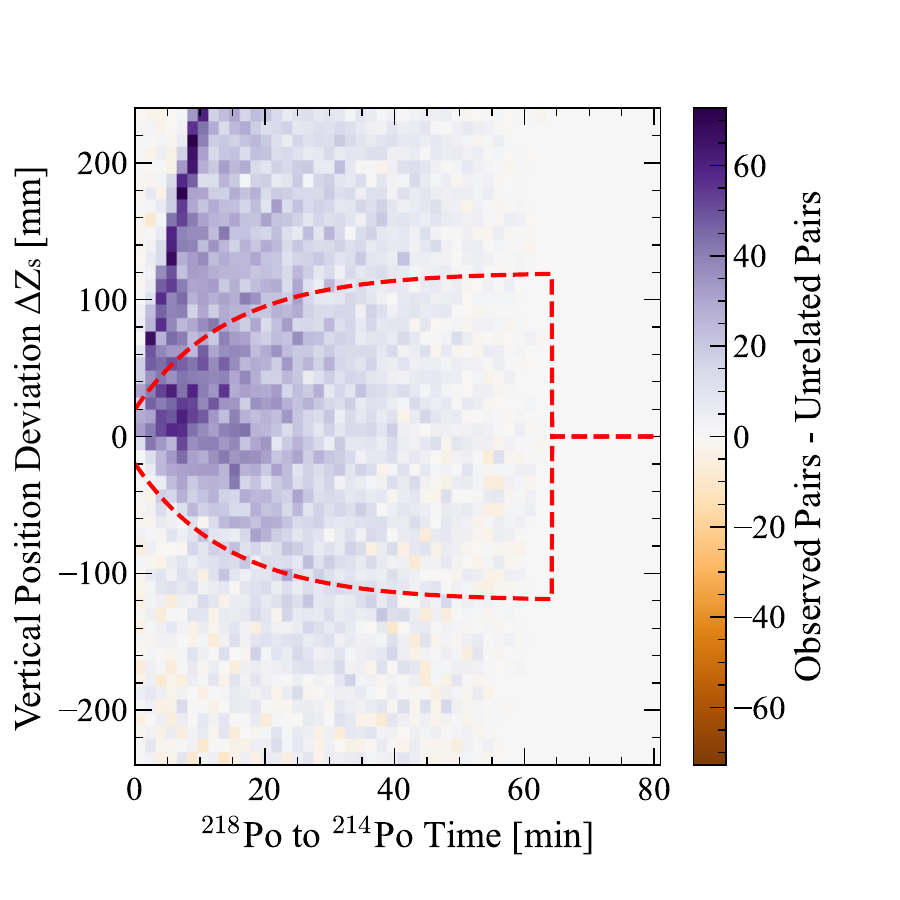}
    } \\ 
    \begin{tabular}{C{\columnwidth} C{\columnwidth} } 
    A & B
    \end{tabular}
    \subfloat{
    \includegraphics[width=\columnwidth]{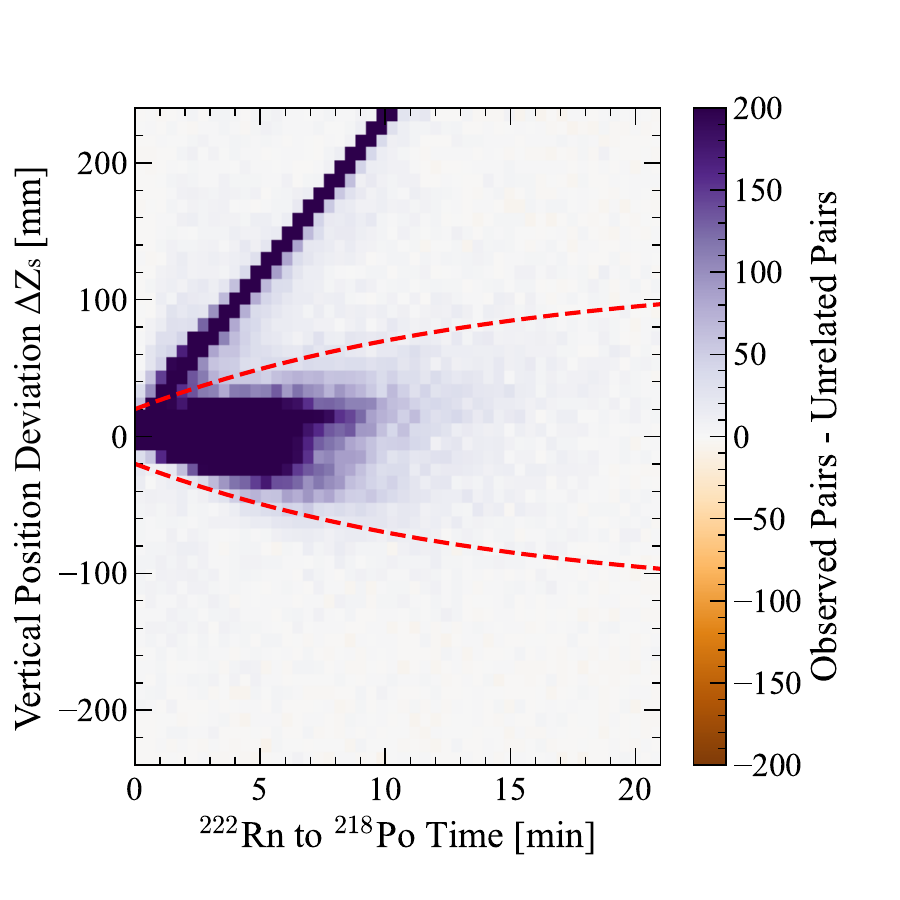}
    } \\ 
    \begin{tabular}{C{\columnwidth} } 
    C
    \end{tabular}
    \caption{
     Charged progeny search performance for \PoBiPo (panels A-B) and \RnPo (panel C) pairs, showing horizontal (panel A) and vertical (panels B and C) spatial deviations between observed progeny event locations and the locations projected by charged streamlines extended from progenitor decays. Red dashed lines indicate the selection window defined in Eq.~(\ref{EQ:StreamCut},\ref{EQ:StreamParamsCharged}), and panels A and B have applied the selection in the complementary coordinate.  All panels have statistically subtracted the unrelated pair background, determined from a 24-hour-shifted search. Because panel B shows no clear separation between the charged population ($\Delta Z_\mathrm{s}=0$) and the semi-charged population ($\Delta Z_\mathrm{s}>0$ but below the prominent neutral population) the vertical window around charged streamlines is tuned on \RnPo (panel C) and projected forward in $\Delta T$ to the maximum ion drift time of 64 minutes, corresponding to ions traversing the full TPC. 
    As in Figure~\ref{fig:Paraview_Neutral_Tuning}, uncertainties in the background-subtracted rate can be estimated from the color-spread in the regions where the average pair excess is zero, with typical per-bin statistical uncertainties of 11, 4, and 9 pairs in panels A, B, and C, respectively.
    }
    \label{fig:Paraview_Charge_Tuning}
\end{figure*}

Figure~\ref{fig:Paraview_Neutral_Tuning} illustrates this process, showing observed \PoBiPo deviations from neutral streamlines and the corresponding selection windows.  These windows extend to $\Delta T$=81~min and are parameterized as
\begin{equation}    \label{EQ:StreamCut}
    \begin{aligned}
        \Delta XY_\mathrm{s} &< g + h(1-\exp(-\Delta T/\kappa)) \\ 
        |\Delta Z_\mathrm{s}| &< i + j(1-\exp(-\Delta T/\lambda)),
    \end{aligned}
\end{equation}
where the parameters for the neutral case are:  
\begin{equation}  
\label{EQ:StreamParamsNeutral}
    \begin{aligned}
        g = 15~\mathrm{mm}, \phantom{111} h &= 40~\mathrm{mm}, \phantom{111} \kappa = 800~\mathrm{s}, \\ 
        i = 30~\mathrm{mm}, \phantom{111} j &= 30~\mathrm{mm}, \phantom{111} \lambda = 800~\mathrm{s}.
    \end{aligned}
\end{equation}
As was done for the \RnPo selection windows, the choice of functional form and setting of parameters is done by hand.

Selection windows around charged streamlines are complicated by the fact that \PoBiPo pairs span the intermediate $^{214}$Pb decay. This is already evident in panel B of Figure~\ref{fig:Paraview_Neutral_Tuning} from the presence of ``semi-charged'' \PoBiPo pairs where either the $^{214}$Pb is charged and $^{214}$Bi is not, or vice versa.  This leads to a population of true \PoBiPo pairs that lie between the projected neutral and charged streamlines.  For the purpose of defining selection windows to be used in \PoPb tagging, the semi-charged population is not of interest, but even for fully-charged \PoBiPo pairs the unknown time of the intervening $^{214}$Pb decay adds uncertainty, as according to Eq.~(\ref{EQ:InstantaneousIonDriftVelocity}) the freshly created $^{214}$Bi$^+$ ion can be expected to move faster just after the unknown time of its creation. For simplicity, we ignore this effect; \emph{i.e.}, our charged streamlines assume the fully-charged case is well modeled by treating the $^{214}$Pb$^+\!\rightarrow\,\!^{214}$Bi$^+$ system as a single continuously existing ion. This underestimates the resulting charge drift by $\sim$2~mm, which is small compared to the selection window. Deviations from these charged streamline predictions are shown in Figure ~\ref{fig:Paraview_Charge_Tuning}.

This approach is sufficient to define the selection window in $\Delta XY_\mathrm{s}$, but in $\Delta Z_\mathrm{s}$ there is no clear distinction between fully-charged and semi-charged pairs.  Instead, \RnPo pairs are used to inform the $\Delta Z_\mathrm{s}$ selection window around charged streamlines (panel C of Figure~\ref{fig:Paraview_Charge_Tuning}), despite the \RnPo population extending only to $\Delta T$ of $\sim$15~min.  
The resulting selection parameters around charged streamlines, using the parameterization in Eq.~(\ref{EQ:StreamCut}) and extending to $\Delta T$=64~min (the maximum drift time for ions traversing the full TPC), are
\begin{equation}
\label{EQ:StreamParamsCharged}
    \begin{aligned}
    g^\textbf{\plus} = 15~\mathrm{ mm}, \phantom{11} h^\textbf{\plus} &= 30~\mathrm{ mm}, \phantom{11} \kappa^\textbf{\plus} = 144~\mathrm{s}, \\ 
    i^\textbf{\plus} = 20~\mathrm{ mm}, \phantom{11} j^\textbf{\plus} &= 100~\mathrm{ mm}, \phantom{11} \lambda^\textbf{\plus} = 600~\mathrm{s}.
    \end{aligned}
\end{equation}

\subsection{Evaluation of \PoPb Tag Performance}
\label{subsec:PoPb}
The tag windows defined using \PoBiPo pairs can now be used in a \PoPb search to identify likely $^{214}$Pb decay events.  Unlike the previous pair searches, the sets of candidate progeny events considered here are neither pure (there is no \emph{a priori} identification of $^{214}$Pb events) nor complete (this analysis considers only single-scatter events, while most $^{214}$Pb decays include gamma emission and are reconstructed as multi-scatter events).

Two sets of candidate $^{214}$Pb decays are considered: a $^{214}$Pb-dominated, low-statistics sample from 18--75~keV$_\textrm{ee}$ (reconstructed ER-equivalent energy~\cite{LZ-backgrounds}), and a high-statistics, high-background sample from 10--1,000~keV$_\textrm{ee}$.  Both samples consider only single-scatter events (cuts defined in \cite{LZ-firstResults}) in LZ's 5.5~t fiducial volume. The $^{214}$Pb-dominated sample also includes an ER-band cut, selecting events within 5-$\sigma$ of the beta-decay band as simulated by the LZ-tune of NEST~\cite{LZ-firstResults,NEST1,NEST2}.

For each sample we consider three tag performance metrics, summarized in Table~\ref{tab:PerformanceMetrics}. These include the fraction of exposure contained in the tag windows ($f_{\mathrm{exp}}$); the fraction of $^{214}$Pb decays contained in those tag windows ($\epsilon_{\mathrm{tot}}$); and the fraction of $^{214}$Pb decays contained in the windows extended from their progenitor $^{218}$Po ($\epsilon_{\mathrm{pair}}$). The latter two metrics are related but subtly different: $\epsilon_{\mathrm{tot}}$ is the relevant background reduction metric for rare event searches and includes $^{214}$Pb tagged by unrelated pairs, while $\epsilon_{\mathrm{pair}}$ reflects the efficiency for true progenitor-progeny pairing only. These metrics are related by 
\begin{equation}
\epsilon_{\mathrm{tot}} \gtrsim \epsilon_{\mathrm{pair}} + (1-\epsilon_{\mathrm{pair}})f_{\mathrm{exp}},
\label{EQ:PairRateComps}
\end{equation} 
where the relation is a one-sided approximation because the $^{214}$Pb population is not uniform in the detector, and the tagging windows, which by construction target this population, show the same non-uniformity.  This results in an unrelated $^{214}$Pb tagging rate that in principle may be slightly greater than $f_{\mathrm{exp}}$, though in practice the effect is smaller than the systematic uncertainties on $\epsilon_{\mathrm{tot}}$ and $\epsilon_{\mathrm{pair}}$
discussed below.

Two methods for determining tagging efficiency are presented: a spectral fit that directly measures $\epsilon_{\mathrm{tot}}$, and a measurement based on the pair excess over a time-shifted search, directly measuring $\epsilon_{\mathrm{pair}}$.  Because $\epsilon_{\mathrm{tot}}$ is the parameter needed for rare-event analyses (\emph{e.g.}, in the 2024 LZ WIMP search \cite{LZ-WS2024}, $\epsilon_{\mathrm{tot}}$ is a constrained floating parameter in the background model), the result of the former technique, which gives $\epsilon_{\mathrm{tot}}$ without reliance on Eq.~(\ref{EQ:PairRateComps}), is the one quoted in the abstract and conclusions of this work.

\begin{table*}[!tb]
    \centering
    {\renewcommand{\arraystretch}{1.2}
    \begin{tabular}{|c|c|c|c|c|c|c|c|}
    \hline
    Tag & Evaluation & Energy & $^{214}$Pb $\beta$-decay& $f_{\mathrm{exp}}$ & $f_{\mathrm{exp}}$ & $\epsilon_{\mathrm{pair}}$ & $\epsilon_{\mathrm{tot}}$ \\
    Window & Method & Range & Final state & (direct) & (proxy) & \multicolumn{2}{c|}{\boldmath{XX$\%\pm$(stat)$\pm$(sys)}} \\
    \hline
    \hline
    Cylinder: Eq.~(\ref{EQ:StreamCut}) & Spectral fit & \multirow{2}{*}{18--75~keV} & \multirow{3}{*}{Ground} & \boldmath{$9.0\%$} & $9.3\%$ & $59\%$ & \boldmath{ $63\%\pm6\% \pm 7\% $} \\
    \cline{1-2}\cline{5-8}
    \multirow{3}{*}{Voxelized} & \multirow{3}{*}{Shifted search} & & & \multirow{3}{*}{\boldmath{$15.6\%$}} & \multirow{3}{*}{$17.0\%$} & \boldmath{$61\%\pm6\%\pm7\%$} & $67\%$ \\
    \cline{3-3}\cline{7-8}
    & & \multirow{2}{*}{10--1,000~keV} & & & & \boldmath{$54\%\pm3\%\pm7\%$} & $61\%$ \\
    \cline{4-4}\cline{7-8}
    & & & Excited & & & \boldmath{$63\%\pm1\%\pm6\%$} & $69\%$ \\
    \hline
    \end{tabular}
    }

    \caption{Summary of $^{214}$Pb reduction metrics for the 5.5 tonne fiducial volume, including the fraction of exposure contained in the tag windows ($f_{\mathrm{exp}}$); the fraction of $^{214}$Pb decays tagged by their progenitor $^{218}$Po ($\epsilon_{\mathrm{pair}}$); and the fraction of of $^{214}$Pb decays contained in any tag window ($\epsilon_{\mathrm{tot}}$).  Two values of $f_{\mathrm{exp}}$ are given for each tag construction: one from direct integration of the tagged volume, and one by measuring the tagging rate in a proxy population for which only accidental tagging is possible, as described in the text.  The accuracy of the proxy measurement of $f_{\mathrm{exp}}$ is limited by the non-uniformity of the proxy population, and in both cases is consistent with the direct measurement.  
    Each row includes a direct measurement of \emph{either} $\epsilon_{\mathrm{pair}}$ (by statistically subtracting a time-shifted search) or $\epsilon_{\mathrm{tot}}$ (by fitting energy spectra of tagged and untagged populations), with systematic uncertainties as described in the text.  The complimentary $\epsilon$ is estimated (without uncertainty) using Eq.~(\ref{EQ:PairRateComps}).}
    \label{tab:PerformanceMetrics}
\end{table*}

\begin{figure*}[tb]
    \centering
    \includegraphics[width=\columnwidth]{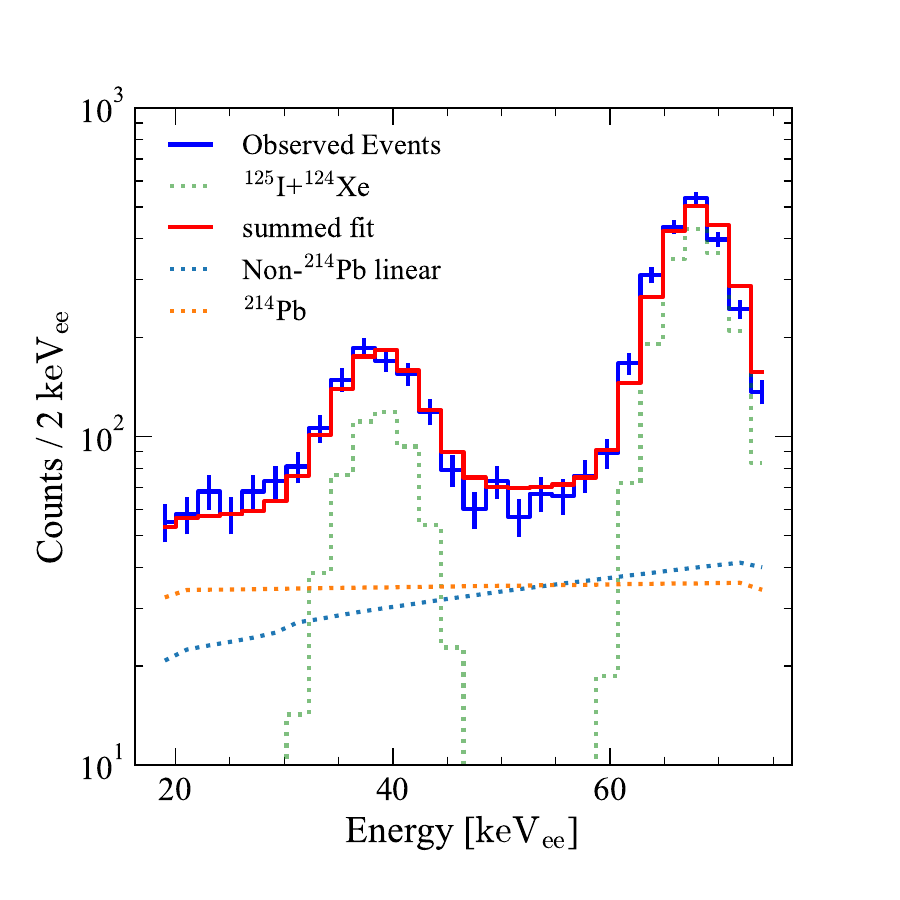}
    \includegraphics[width=\columnwidth]{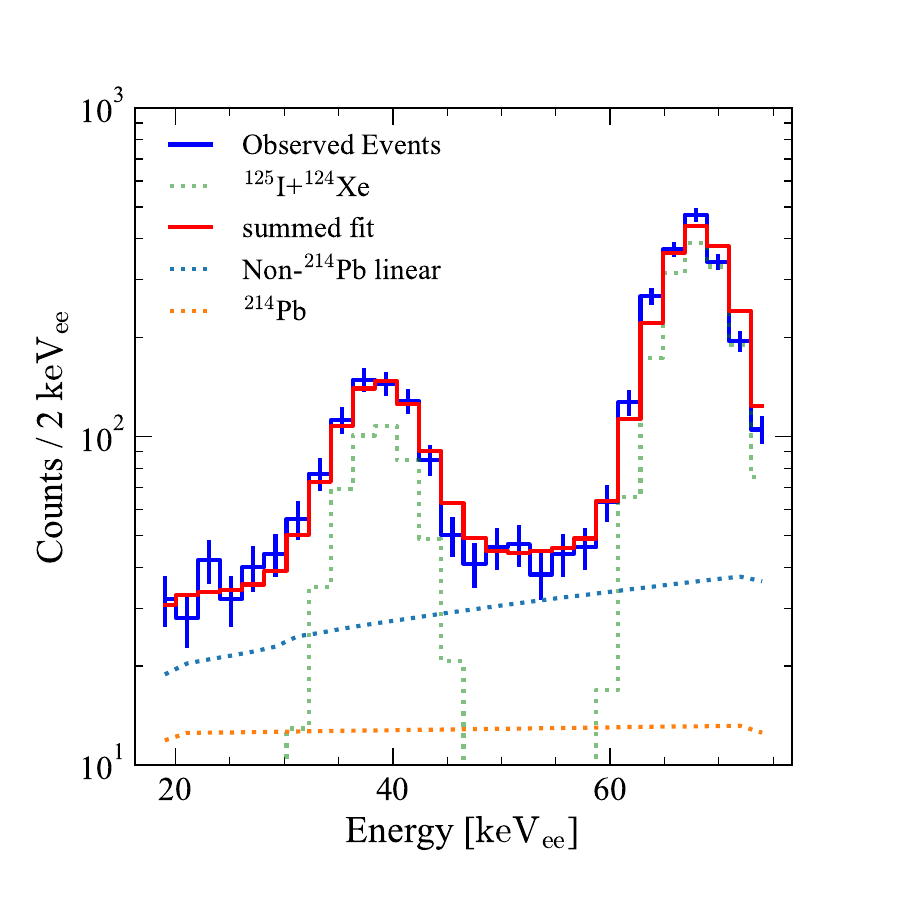}
    \caption{Single-scatter background spectra in the 5.5 tonne fiducial volume, showing the $^{214}$Pb-dominated region from 18--75~keV$_\textrm{ee}$ (reconstructed electron-equivalent energy) before (left) and after (right) removing events selected by the $^{214}$Pb tag defined in Eq.~(\ref{EQ:StreamCut}--\ref{EQ:StreamParamsCharged}). The blue histogram in each plot shows the measured spectrum with statistical error-bars, while the red histogram in each plot shows the best-fit of a background model composed of two Gaussian peaks (green dashed line) with fixed means for K- and L-shell electron-capture decays of $^{125}$I (plus a small rate of unresolved $^{124}$Xe double-electron-capture decays at nearly the same energies), plus a roughly linear spectrum that includes $^{214}$Pb (orange dashed line) and other continuous ER contributions (blue dashed line, including $^{85}$Kr, $^{136}$Xe, $^{212}$Pb, and solar neutrino-electron scattering). The decomposition of the continuous spectrum into $^{214}$Pb and non-$^{214}$Pb components is taken from \cite{LZ-backgrounds}.  After removing tagged events the spectrum is re-fit with two free parameters, giving reduction factors for $^{214}$Pb and non-$^{214}$Pb components. 
    These fits indicate that non-$^{214}$Pb components are reduced by $f_{\mathrm{exp}}=9.3\%\pm0.6\%\textrm{(stat)}$, while the $^{214}$Pb rate is reduced by $\epsilon_{\mathrm{tot}}=63\%\pm6\%\textrm{(stat)}\pm7\%\textrm{(sys)}$.
    }
    \label{fig:EnergyFit}
\end{figure*}

\subsubsection{Low-energy spectral analysis of $\epsilon_{\mathrm{tot}}$}

Figure~\ref{fig:EnergyFit} shows the effect of the $^{214}$Pb tag on the $^{214}$Pb-dominated sample.  The energy spectrum in this region includes the roughly flat, continuous electron-recoil background described in \cite{LZ-backgrounds} (a combination of $^{214}$Pb, $^{85}$Kr, $^{136}$Xe, $^{212}$Pb, and solar neutrino-electron scattering),
plus two peaks at 40.4 and 67.3~keV from $^{125}$I electron-capture decay (with small unresolved contributions from $^{124}$Xe double-electron-capture at 37 and 64~keV \cite{LZ-Xe124}).
The observed spectrum prior to applying the $^{214}$Pb tag is fit to a sum of four components: the $^{214}$Pb and non-$^{214}$Pb continuous spectra with shapes and relative normalization taken from \cite{LZ-backgrounds} (with the overall normalization a free parameter), and two Gaussians with means fixed at the known energies (with each amplitude and width a free parameter).  Tagged events are then removed and the spectrum is re-fit, keeping the previous best-fit spectrum fixed and fitting the tagged spectrum with two free parameters, $\epsilon_{\mathrm{tot}}$ and $f_{\mathrm{exp}}$, the former reducing the $^{214}$Pb component of the spectrum and the latter reducing all other components (taking the approximation that those components are uniformly distributed in the detector).

The fits indicate $f_{\mathrm{exp}}=9.3\%\pm0.6\%\textrm{(stat)}$, consistent with the $f_{\mathrm{exp}}=9.0\%$ found by direct integration of the exposure contained in the tagged volume.
Meanwhile the tag reduces the best-fit $^{214}$Pb contribution from 976.9 events to 358.2 events, giving $\epsilon_{\mathrm{tot}}= 63\%\pm6\%\textrm{(stat)}\pm7\%\textrm{(sys)}$. 
The systematic uncertainty in  $\epsilon_{\mathrm{tot}}$ reflects uncertainty in the $^{214}$Pb content of the linear background (\emph{i.e.}, the prior for the pre-tag fit) as given in \cite{LZ-backgrounds}, which in turn comes predominantly from uncertainty in the $^{214}$Pb branching ratio to the $^{214}$Bi ground state \cite{Pb214_ENDSF}.  
The fraction of $^{214}$Pb decays correctly tagged by their progenitor is estimated using Eq.~(\ref{EQ:PairRateComps}) to be $\epsilon_{\mathrm{pair}}\lesssim59\%$.

\subsubsection{Broad-spectrum analysis of $\epsilon_{\mathrm{pair}}$}
\label{subsubsec:voxeltag}

\begin{figure*}[p]
    \centering
    \includegraphics[width=2\columnwidth]{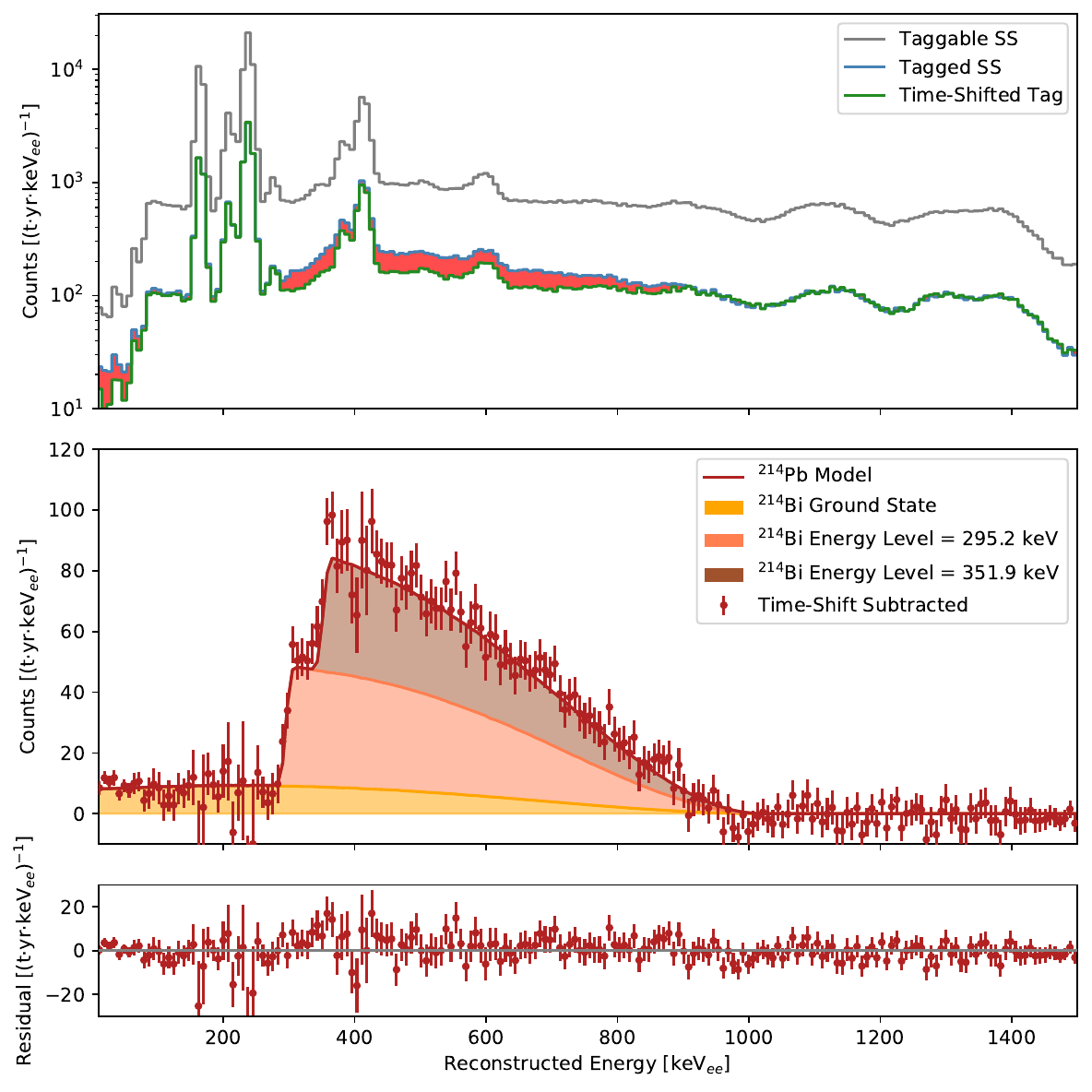}
    \caption{Reconstructed energy spectra for single scatter interactions in the 5.5 tonne fiducial volume.  The top panel shows spectra for: all single-scatter events in the taggable data (gray); events tagged as potential $^{214}$Pb (blue, top boundary of red shaded region); and events tagged by time-shifted searches, here the average tag rate from searches shifted by 90 and 180 minutes (green, bottom boundary of red shaded region).  The center panel shows the excess tag rate (unshifted tag spectrum minus shifted tag spectrum) corresponding to the shaded region in the top panel, overlaid on the $^{214}$Pb single-scatter background model used in Ref.~\cite{LZ-backgrounds}.  The decay-to-ground-state (yellow) and decay-to-excited-state (sum of orange and brown) portions of the model are normalized separately, as described in the text.  Absolute residuals between the subtracted spectrum and the normalized $^{214}$Pb model are shown in the bottom panel. Statistically significant residuals appear in regions of the spectrum dominated by peaked backgrounds (primarily $^{131\mathrm{m}}$Xe, $^{127}$Xe, $^{129\mathrm{m}}$Xe, and $^{125}$Xe~\cite{LZ-backgrounds}), likely due to temporal variation in peak position and height leading to systematic errors in the subtraction.
    }
    \label{fig:EnergyFitFull}
\end{figure*}

To simplify the application of the $^{214}$Pb tag to larger samples, such as the full single-scatter spectrum, the LXe-TPC is divided into 1.5-cm cubic voxels and a lookup table is created listing the time windows for which each voxel is `tagged''.  A voxel is flagged as ``tagged'' whenever it overlaps with any cylindrical exclusion region defined by Eq.~(\ref{EQ:StreamCut}--\ref{EQ:StreamParamsCharged}), resulting in a voxelized tag volume that fully covers the previously defined tag volume, increasing $f_{\mathrm{exp}}$ to $15.6\%$ as determined from direct tag volume integration.  The effect of this voxelized tag on the full single-scatter spectrum is shown in Figure~\ref{fig:EnergyFitFull}.  Although $^{214}$Pb is a subdominant background above 80~keV, statistical subtraction of a time-shifted search recovers the expected $^{214}$Pb spectrum Q-value and gamma/internal-conversion shoulders from decays to excited states.  The correctly-tagged $^{214}$Pb fraction $\epsilon_{\mathrm{pair}}$ can be determined by comparing the subtracted spectrum to the $^{214}$Pb background model in \cite{LZ-backgrounds}.

The evaluation of $\epsilon_{\mathrm{pair}}$ is done separately for decays to ground state and decays to excited state due to the independent systematic uncertainties in the background model in these two cases:  the uncertainty on decays to the ground state is dominated by uncertainty in the ground state branching ratio, while uncertainty on decays to excited states includes uncertainty on the fraction of gamma-emitting decays reconstructed as single-scatter events due to the short range of the gamma ray.  Such events make up roughly 20\% of the measured single-scatter rate above 300~keV$_\textrm{ee}$, with the balance coming from decays to ground state or to excited state with internal conversion rather than gamma ray emission.  This analysis finds that $\epsilon_{\mathrm{pair}}=54.0\%\pm2.6\%\textrm{(stat)}\pm7\%\textrm{(sys)}$ for $^{214}$Pb decays to ground state, while $\epsilon_{\mathrm{pair}}=63.4\%\pm0.8\%\textrm{(stat)}\pm6\%\textrm{(sys)}$ for $^{214}$Pb decays to excited states.  If restricted to the Pb-dominated 18--75 keV$_\textrm{ee}$ window (includes decays to ground state only), the time-shift-subtraction analysis finds $\epsilon_{\mathrm{pair}}=60.6\%\pm5.6\%\textrm{(stat)}\pm7\%\textrm{(sys)}$, statistically consistent with the 59\% found above.  Meanwhile, 17.0\% of all fiducial volume events (which are not spatially uniform) are caught by the shifted tag, qualitatively consistent with the calculated $f_{\mathrm{exp}}=15.6\%$.  This style of voxelized tag was implemented with similar performance in LZ's 2024 WIMP search~\cite{LZ-WS2024}. 

\subsection{Modeling \PoPb Tag Performance} \label{subsec:chargemodel}
It is useful to compare the observed tag performance to the expected performance based on known sources of inefficiency, including missed $^{218}$Po progenitors, $^{214}$Pb surviving past the 81-minute search window, and neutralization of $^{214}$Pb ions leading to $^{214}$Pb decays that fall between the charged and neutral streamline selection windows.  Modeling these effects requires values for the charge branching fraction of $^{218}$Po decays and the mean neutralization time $\tau_\mathrm{n}$ for drifting $^{214}$Pb$^+$ ions.  Measurements of these quantities are described in Appendix~\ref{sec:Supplemental} and summarized in Table~\ref{tab:ChargeBranching}.

\begin{table}[t]
    \centering
    \setlength{\extrarowheight}{2pt}  
    \begin{tabular}{|c|c|c|}
        \hline
         Decay & Charge Branching & Charge Neutralization  \\
        Parent-Progeny & Fraction & Time $\tau_\mathrm{n}$ [min] \\ 
        \hline
        \RnPo & 0.49 $\pm$ 0.01 & 49 $\pm$ 4 \\ 
        \PoPb & 0.48 $\pm$ 0.12 & - \\ 
        \PbBi & 0.74 $\pm$ 0.05 & - \\
       \hline
    \end{tabular}
    \caption{Best fit charge branching fractions for each radon chain decay, as derived from the performance of the corresponding neutral streamline searches. 
    Progeny charge state is assumed to be independent of progenitor charge state, and charged-progeny neutralization is assumed to be the same for all species, measured only for \RnPo pairs. 
 Charge branching fractions are consistent with those previously measured in \cite{EXO-200chargedions}.}
    \label{tab:ChargeBranching}
\end{table}

It is now possible to construct a simple Monte-Carlo simulation for charged ion transport and evolution, to estimate the effects of cathode plateout and ion neutralization on tagging efficiency. The simulation draws initial progeny $Z$ positions from observed progenitor decay locations, initial progeny charge state using the charge branching fractions in Tab.~\ref{tab:ChargeBranching} (assumed binomial draws, independent of progenitor charge state), and progeny neutralization and decay times from exponential distributions using the $\tau_n$ in Tab.~\ref{tab:ChargeBranching} and the known half-lives, respectively.  The charged ion drift velocities described by Eq. (\ref{EQ:InstantaneousIonDriftVelocity}) are applied while the progeny is charged, and if an ion reaches $Z$=0 it is assumed to plate out on the cathode.

We can use this model to predict the performance of a \PoPb tag, assuming a perfect transport model for both neutral and charged progeny with tag volumes as defined above. Taking the known \UChPoE~spatial distribution as input, the simulation predicts a tagging efficiency of 75\%$\pm$1\% for $^{214}$Pb decays in the 5.5~t fiducial volume. Contributing to the modeled inefficiency are the \UChPoE~selection efficiency (95.9\% DAQ livetime) and the 3~half-life tag window (87.5\% efficiency for neutral progeny, no effect on charged progeny since they are swept from the fiducial volume in less than 3~half-lives). The remaining modeled inefficiency is due to neutralized charged progeny decaying between the neutral and charged tag windows. The uncertainty on the simulated efficiency stems from the uncertainty on the charge neutralization time.   
The gap between the observed (63\%) and expected (75\%) tagging efficiency suggests that some room for improvement in the $^{214}$Pb transport model remains.  The \PoBiPo tag shows a similar gap between observed and modeled performance, with an observed pairing rate $\sim$0.8$\times$ the modeled rate.

\begin{figure*}[t]
\centering
    \subfloat{
    \includegraphics[width=\columnwidth]{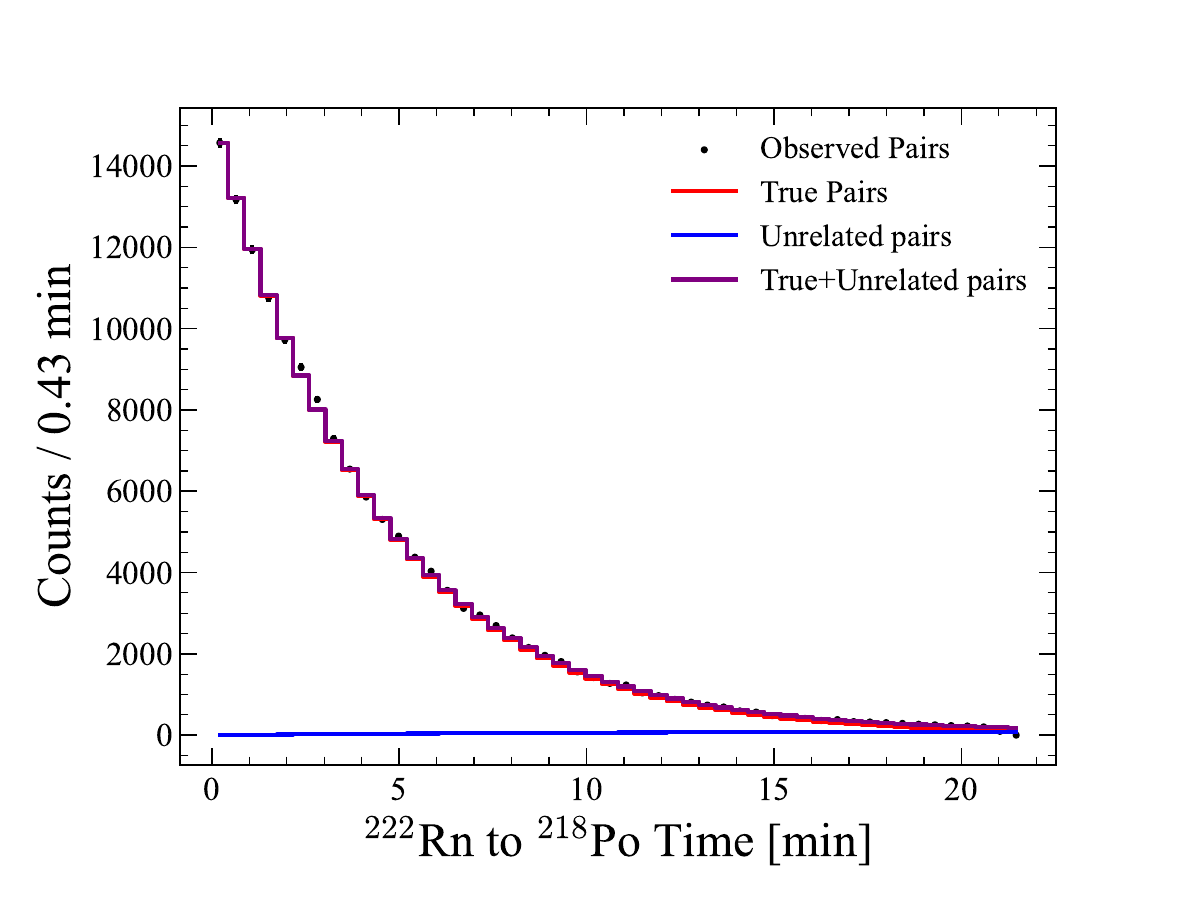}
    \includegraphics[width=\columnwidth]{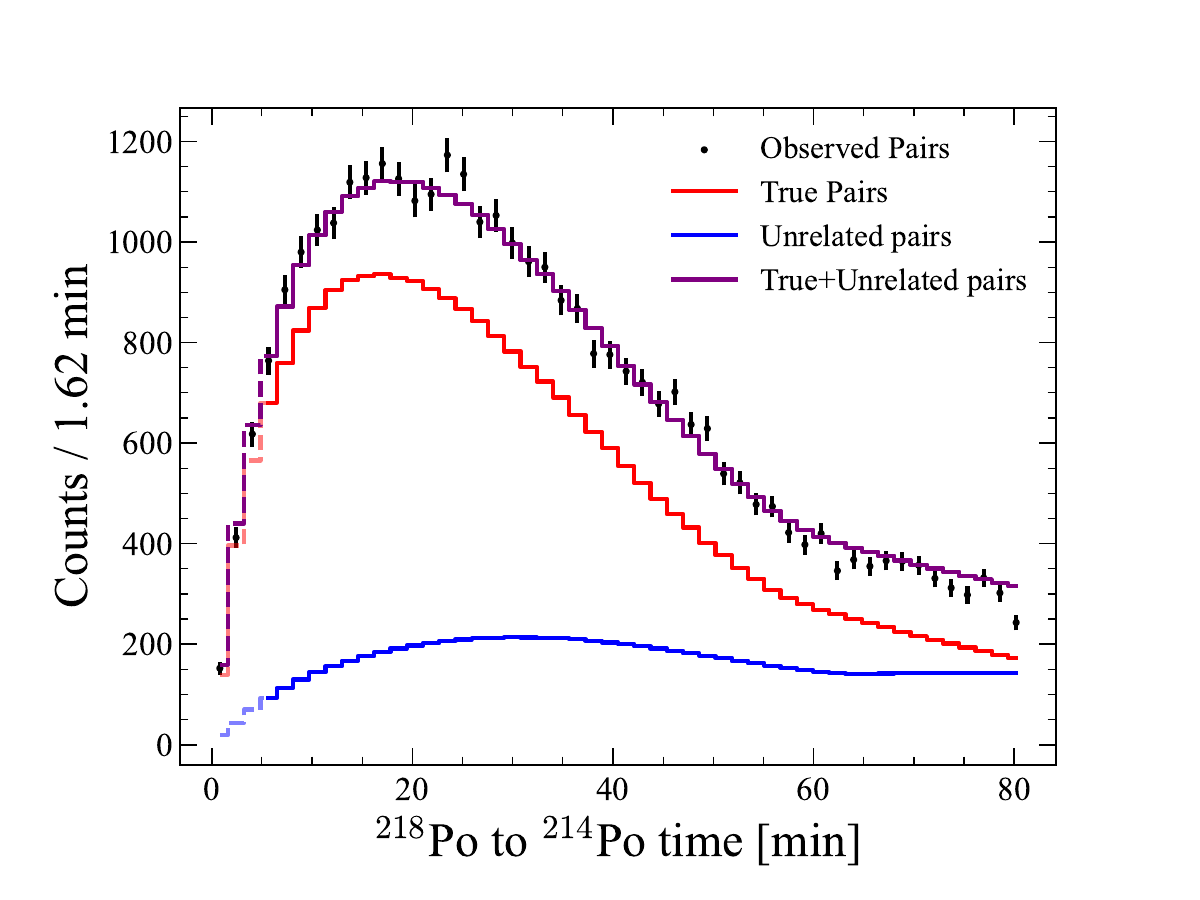}
    } \\
    \begin{tabular}{C{\columnwidth} C{\columnwidth} } 
    A & B
    \end{tabular}
    \subfloat{
    \includegraphics[width=\columnwidth]{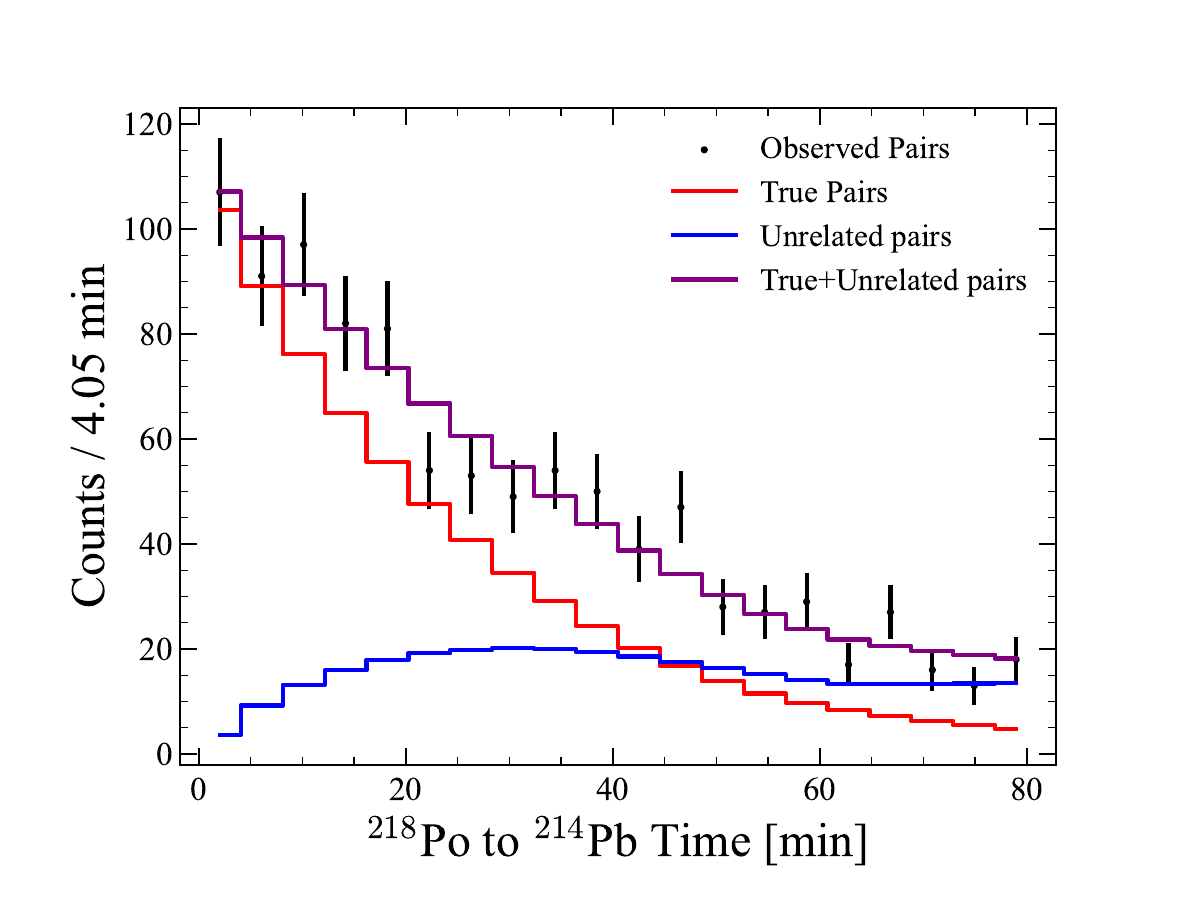}
    }\\
    \begin{tabular}{C{\columnwidth} } 
    C
    \end{tabular}
    \caption{Distributions of time separations ($\Delta T$) for decay pairs tagged by the selection volume defined in Eq.~(\ref{EQ:StreamCut}--\ref{EQ:StreamParamsCharged}), showing: \RnPo (panel A), \PoBiPo (panel B), and \PoPb (panel C, taking the 18--75~keV$_\textrm{ee}$ $^{214}$Pb-dominated sample).  Only pairs where the progeny decays in the 5.5~t fiducial volume are included.  Black points show observed pair rates with statistical error bars, without subtracting the unrelated pair background. Lines show the modeled $\Delta T$ distribution (purple) broken into correctly-paired decays (red) and accidental parings of unrelated decays (blue). 
 The true pair distributions are constructed using a Monte-Carlo simulation that takes observed progeny starting locations as input and considers progeny mean lifetimes (including the two-step decay for ${}^{218}$Po--${}^{214}$Pb--${}^{214}$BiPo in panel B) and expectations for charged progeny drift and neutralization.  
 The unrelated pair distributions are given by time-shifted searches, with shapes reflecting both the growth of the selection volume over time and the reduction of the selection volume from the truncation of charged streamlines leaving the fiducial volume. The normalizations of the two pair types are allowed to float, with a Poisson prior on the unrelated pair rate.  The solid (dashed) portions of model lines in panel B indicate the intervals in $\Delta T$ included (excluded) in the fit.
    \label{fig:ParaViewLifetimeFits}
    }
\end{figure*}

Further insight can be gained from the modeled $\Delta T$ distributions for identified pairs.  Figure~\ref{fig:ParaViewLifetimeFits} fits the observed $\Delta T$ distributions for \RnPonospace, \PoPbnospace, and \PoBiPo fiducial-volume searches to a combination of unrelated pairings (constrained by shifted searches) and correct pairings (distribution shape given by Monte-Carlo simulation, with floating normalization).
The true-pair template reflects the known half-lives of the relevant decays combined with a slightly faster fall off for charged pairs due to ions drifting out of the fiducial volume or plating out on detector surfaces. 
The unrelated pair component normalization is constrained by a 24-hour-shifted search with a simple Poisson prior, while the shape reflects the combined effects of acceptance volume growth described by Eq. (\ref{EQ:StreamCut}), as well as charged acceptance windows drifting out of the fiducial volume.

In all three pair searches, the shape of the observed $\Delta T$ distribution is roughly consistent with expectations.
A small excess in the \RnPo pairing rate is observed for $\Delta T$ of 2~to 3~minutes. This corresponds to the time it takes charged progeny to drift from the liquid surface to the fiducial volume, suggesting imperfect modeling of the progenitor $^{222}$Rn population and/or $^{218}$Po$^+$ transport in and around the extraction region.
There is also a $\sim$10\% bigger deficit of observed \PoBiPo pairs at $\Delta T \gtrsim 60$~minutes, relative to the model.  A likely source of this discrepancy is also in the charged progeny branch, where the charge drift model and width of the $\Delta Z_\mathrm{s}$ window are based on \RnPo pairs, measuring only $^{218}$Po$^+$ ion drift and extending only to $\Delta T$ of 15 minutes, but then applied to  \PoPb and \PoBiPo searches that extend to 81 minutes and are sensitive to both $^{214}$Pb$^+$ and $^{214}$Bi$^+$ ion drift.
Neither of these discrepancies affect the reported measurements of $^{214}$Pb tagging efficiency, but they do indicate avenues for further study and provide hints for how tagging efficiency might be improved.

One likely method to improve the efficiency of the $^{214}$Pb tag would be to include streamlines projected backward from $^{214}$BiPo events.  This would allow the use of neutral streamlines to tag semi-charged decay paths with a charged $^{214}$Pb$^+$ (the presumed source of the current disagreement between the modeled and observed tag efficiency) but neutral $^{214}$Bi.  In addition to improving agreement between model and observation, this approach would increase the modeled efficiency by reducing the impact of ion neutralization, as $^{214}$Pb events would now be lost to neutralization only when both the $^{214}$Pb and $^{214}$Bi ions neutralize.  Such a tagging scheme is currently under development, and will be part of a more sophisticated analysis where the binary tag is replaced by a continuous $^{214}$Pb score calculated for each candidate event without voxelization.

\section{Impact on future dark matter searches}
\label{sec:Impact}
A high-efficiency tag for $^{214}$Pb impacts dark matter discovery in two complementary ways:  it significantly reduces the leading source of low energy ER backgrounds, and it also provides for the first time a low-energy ER calibration dataset taken concurrently with science data.  Mitigation of the $^{214}$Pb background is demonstrated in~\cite{LZ-WS2024}, using a voxelized binary tag (as described in Section~\ref{subsubsec:voxeltag}) to partition LZ's 2024 WIMP search into  a $^{214}$Pb-rich exposure containing events in tagged voxels and a complementary $^{214}$Pb-poor exposure where the residual $^{214}$Pb rate makes up less than half of the observed low energy ER background.  The 2024 WIMP search combines these two exposures in a single profile likelihood analysis that includes $\epsilon_{tot}$ as a floating parameter.  The impact of the $^{214}$Pb-tag on a given search depends on the overlap between the specific physics signal and the ER band, and due in part to LZ's ER/NR discrimination a negligible impact was seen in LZ's 2024 WIMP search~\cite{LZ-WS2024}.  The impact of the $^{214}$Pb tag on rare-event searches in the electron-recoil band is more significant, improving the sensitivity of the searches in \cite{LZ:2025zpw} by 5---10\%, depending on the spectral shape of the signal in question.

\begin{figure}[!tb]
\centering
    \includegraphics[width=\columnwidth]{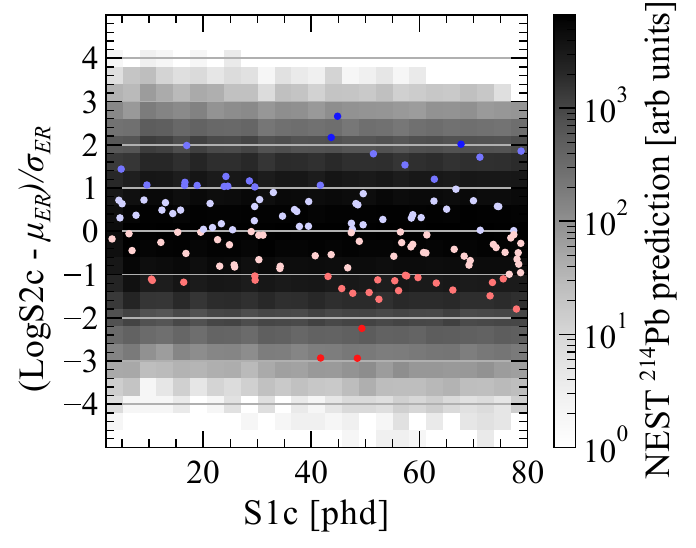}
\caption{WIMP ROI flow-tagged events displayed in S2-distance to the NEST ER band mean versus S1.  The mean $\mu_\mathrm{ER}$ and standard deviation $\sigma_\mathrm{ER}$ of $\log$(S2c) are calculated in slices of S1c.  Point color corresponds to \textit{y}-axis value. NEST expectations (displayed in gray-scale behind the scatter points) come from the NEST $\beta^-$ model with a $^{214}$Pb energy spectrum. Expected and observed event counts in each 1-$\sigma$ band are given in Table~\ref{tab:ER_model_fit}.
} \label{fig:S2comps}
\end{figure}

The impact of the $^{214}$Pb tag as a new calibration dataset is more immediate.  The LZ ER response model (e.g. from NEST \cite{NEST1, NEST2}) is typically tuned with high-rate calibration sources such as $^{212}$Pb and $^{3}$H, which are injected into the LXe-TPC between long periods of science data taking~\cite{LZ-cal}. The $^{214}$Pb-tagged events are unique in being able to provide a high-purity cross validation set that is contemporaneous with science data. Figure~\ref{fig:S2comps} and Table~\ref{tab:ER_model_fit} show the agreement between the NEST ER template tuned on injected $\beta^-$ sources (as used in~\cite{LZ-firstResults,LZ-backgrounds}) and the voxelized flow-tagged events from the LZ WIMP ROI in~\cite{LZ-firstResults}.
We see excellent agreement between the NEST model and the $^{214}$Pb-tagged events, which appear symmetrically across the expected median with a spread in good agreement with expectation.  This in-situ calibration has been especially relevant in LZ's 2024 WIMP search~\cite{LZ-WS2024}, where $^{124}$Xe double-electron-capture decays~\cite{LZ-Xe124} appear as a significant background for the first time. These decays (and other single-electron-capture decays) show reduced charge yield relative to the standard $\beta^-$ ER response~\cite{LZ-ECdecay}, and the absence of events with reduced charge yield in the $^{214}$Pb-tagged population help to confirm the origin of these reduced-charge-yield events.

\begin{table}[tb]
    \centering
    \setlength{\extrarowheight}{2pt}  
    \begin{tabular}{|c|c|c|}
       \hline
       $\sigma$ from ER band  & Expected Counts & Observed Counts  \\
       \hline
        $[-3,-2]$ & 3.2 & 3 \\ 
        $[-2,-1]$ & 18 & 23 \\ 
        $[-1,0]$ & 44 & 45 \\ 
        $[0,1]$ & 45 & 42 \\ 
        $[1,2]$ & 17 & 17 \\
        $[2,3]$ & 2.4 & 3 \\
       \hline
    \end{tabular}
    \caption{Comparison of the NEST $^{214}$Pb band profile to observed event counts from the $^{214}$Pb-tagged sample from the WIMP ROI in~\cite{LZ-firstResults}, in bins of S2-distance to the band mean as shown in Figure~\ref{fig:S2comps}. The NEST expectation is scaled to the observed $^{214}$Pb rate. Each bin is consistent within Poisson fluctuations, and a KS-test comparing the S2 distributions gives a \textit{p}-value of 0.319, indicating consistency between the tagged $^{214}$Pb dataset and the NEST model.}
    \label{tab:ER_model_fit}
\end{table}

\section{Conclusions}
\label{sec:Conclusions}
Alpha-decays in the radon chain have been used to extract information about the fluid flow within the LXe-TPC of the LZ experiment. Identified \RnPo~decay pairs serve as a map of displacement due to fluid flow and charged ion drift, providing the basis for a transport model from which single atom trajectories can be calculated for any charged or neutral radon progeny. This transport model successfully identifies \PoBiPo pairs at time separations up to at least 81 minutes, three times the $^{214}$Pb half-life.  A \UChPb~tag has been developed based on selection volumes defined around predicted trajectories after each progenitor \UChPoE~decay, with a demonstrated reduction in the $^{214}$Pb background rate near the WIMP region of interest of
$\epsilon_{\mathrm{tot}}=63\%\pm6\%\textrm{(stat)}\pm7\%\textrm{(sys)}$,
at a cost of $f_{\mathrm{exp}}=9.0\%$ of live exposure.

Flow tests have demonstrated the capacity for achieving both higher- and lower-mixing states than the flow state used in this analysis. The former underlines the limitations of flow mapping when liquid velocities exceed the ion drift velocity, whereas the latter shows that there is space for further flow optimization. Finally, the \UChPb-tag immediately provides a new low-energy electron recoil calibration set that is contemporaneous to dark matter search data.

\begin{acknowledgments}
\emph{Acknowledgements} - The research supporting this work took place in part at the Sanford Underground Research Facility (SURF) in Lead, South Dakota. Funding for this work is supported by the U.S. Department of Energy, Office of Science, Office of High Energy Physics under Contract Numbers DE-AC02-05CH11231, DE-SC0020216, DE-SC0012704, DE-SC0010010, DE-AC02-07CH11359, DE-SC0015910, DE-SC0014223, DE-SC0010813, DE-SC0009999, DE-NA0003180, DE-SC0011702, DE-SC0010072, DE-SC0006605, DE-SC0008475, DE-SC0019193, DE-FG02-10ER46709, UW PRJ82AJ, DE-SC0013542, DE-AC02-76SF00515, DE-SC0018982, DE-SC0019066, DE-SC0015535, DE-SC0019319, DE-SC0025629, DE-SC0024114, DE-AC52-07NA27344, \& DE-SC0012447. This research was also supported by U.S. National Science Foundation (NSF); the UKRI’s Science \& Technology Facilities Council under award numbers ST/W000490/1, ST/W000482/1, ST/W000636/1, ST/W000466/1, ST/W000628/1, ST/W000555/1, ST/W000547/1, ST/W00058X/1, ST/X508263/1, ST/V506862/1, ST/X508561/1, ST/V507040/1, ST/W507787/1, ST/R003181/1, ST/R003181/2,  ST/W507957/1, ST/X005984/1, ST/X006050/1; Portuguese Foundation for Science and Technology (FCT) under award numbers PTDC/FIS-PAR/2831/2020; the Institute for Basic Science, Korea (budget number IBS-R016-D1); the Swiss National Science Foundation (SNSF)  under award number 10001549. This research was supported by the Australian Government through the Australian Research Council Centre of Excellence for Dark Matter Particle Physics under award number CE200100008. We acknowledge additional support from the UK Science \& Technology Facilities Council (STFC) for PhD studentships and the STFC Boulby Underground Laboratory in the U.K., the GridPP~\cite{faulkner2005gridpp,britton2009gridpp} and IRIS Collaborations, in particular at Imperial College London and additional support by the University College London (UCL) Cosmoparticle Initiative, and the University of Zurich. We acknowledge additional support from the Center for the Fundamental Physics of the Universe, Brown University. K.T. Lesko acknowledges the support of Brasenose College and Oxford University. The LZ Collaboration acknowledges the key contributions of Dr. Sidney Cahn, Yale University, in the production of calibration sources. This research used resources of the National Energy Research Scientific Computing Center, a DOE Office of Science User Facility supported by the Office of Science of the U.S. Department of Energy under Contract No. DE-AC02-05CH11231. We gratefully acknowledge support from GitLab through its GitLab for Education Program. The University of Edinburgh is a charitable body, registered in Scotland, with the registration number SC005336. The assistance of SURF and its personnel in providing physical access and general logistical and technical support is acknowledged. We acknowledge the South Dakota Governor's office, the South Dakota Community Foundation, the South Dakota State University Foundation, and the University of South Dakota Foundation for use of xenon. We also acknowledge the University of Alabama for providing xenon. For the purpose of open access, the authors have applied a Creative Commons Attribution (CC BY) license to any Author Accepted Manuscript version arising from this submission. Finally, we respectfully acknowledge that we are on the traditional land of Indigenous American peoples and honor their rich cultural heritage and enduring contributions. Their deep connection to this land and their resilience and wisdom continue to inspire and enrich our community. We commit to learning from and supporting their effort as original stewards of this land and to preserve their cultures and rights for a more inclusive and sustainable future.

\end{acknowledgments}

\bibliographystyle{apsrev4-1}
\bibliography{references}% Produces the bibliography via BibTeX.
\newpage
\phantom{1}
\newpage
\appendix
\section{Charge branching and ion neutralization}
\label{sec:Supplemental}

The charge branching fractions for the $^{222}$Rn, $^{218}$Po, and $^{214}$Pb decays shown in Table~\ref{tab:ChargeBranching} are determined from the pair excess in the neutral-progeny search windows of the \RnPonospace, \PoPbnospace, and \PoBiPo pair searches, respectively, taking the search definitions in Eq.~(\ref{EQ:StreamCut},\ref{EQ:StreamParamsNeutral}) and considering only pairs with the progenitor decay in the 5.5 tonne fiducial volume and $\Delta T\in[200\,\textrm{s}, 900\,\textrm{s}]$.  The \RnPo pair analysis in Section~\ref{sec:PairFinding} shows the fiducial-volume neutral-progeny tagging in this time window to be consistent with 100\% (flow) efficiency, and the minimum $\Delta T$ of 200~s is sufficient to completely separate the neutral and charged populations.  Basing the charge branching fraction measurement on the fraction of progenitors producing a neutral-progeny tag in this high-efficiency window avoids systematic uncertainties from imperfect modeling of charge plateout and ion drift.

\begin{figure}[tb]
    \centering
    \includegraphics[width=\columnwidth]{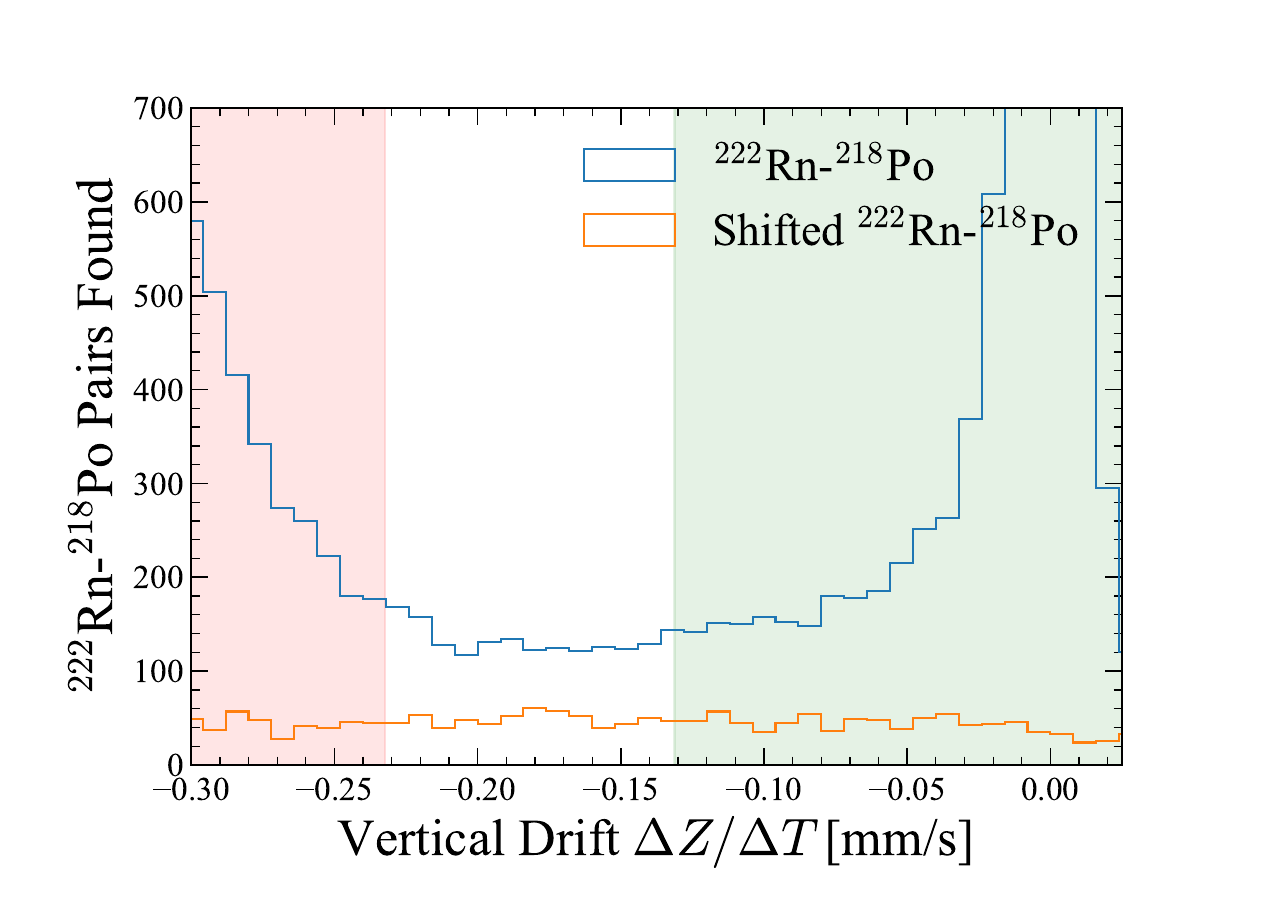}
    \caption{Time-averaged axial velocities ($\Delta Z / \Delta T$) of the \RnPo pairs in the bottom panel of Figure~\ref{fig:IntialRnPoSelections}, marginalized over $\Delta T$ from 200~s -- 900~s.  
    Shaded regions indicate the charged (red) and neutral (green) acceptance volume limits at $\Delta T$=400~s. The clear pair excess in the unshifted search (blue line) over the shifted search (orange line) at velocities between the neutral and charged populations is evidence for the neutralization of charged daughters.}
    \label{fig:IntermediateVZExcess}
\end{figure}

\begin{figure}[tb]
    \centering
    \includegraphics[width=\columnwidth]{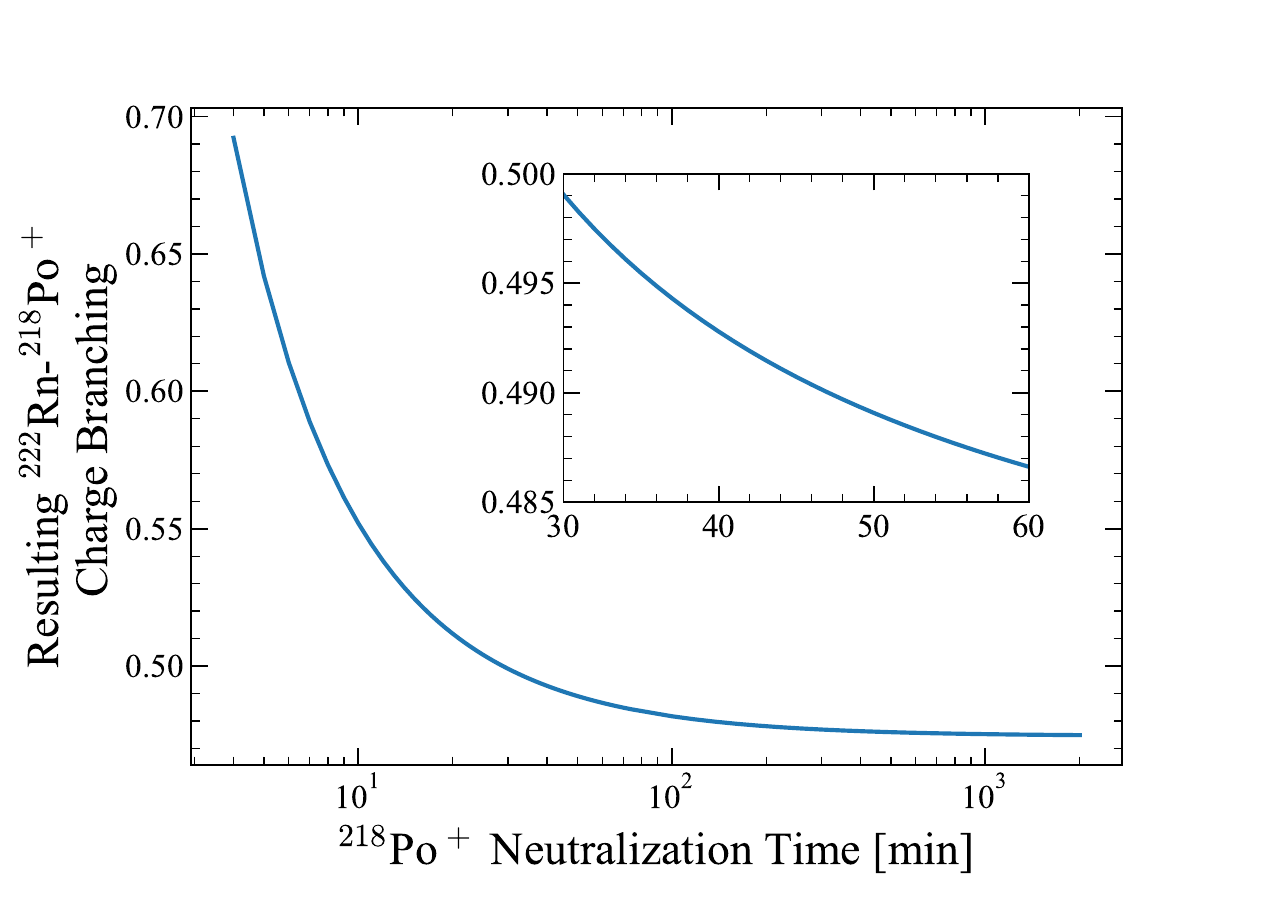}
    \caption{$^{222}$Rn--$^{218}$Po charge branching fractions inferred from a neutral-only pair search restricted to a 200-900 second time window, varying the assumed mean charge neutralization time.  Shorter mean neutralization times result in more charged progeny appearing in the neutral search window,
    and thus require a greater inferred charge branching fraction to produce the same number of pairs tagged in the neutral-progeny search.
    }
    \label{fig:RnPoChargeBranchingForEachNeutralization}
\end{figure}

This measurement is complicated, however, by the neutralization of charged ions, evident in the pair excess seen between the neutral- and charged-daughter populations in Figure~\ref{fig:IntialRnPoSelections} and shown again marginalized over $\Delta T$ in Figure~\ref{fig:IntermediateVZExcess}.  This neutralization occurs on $\mathcal{O}$(10)-minute time scales, short enough that some neutralized charged progeny appear in the neutral progeny search window. This population must be accounted for when using the neutral search window to determine the charge branching fraction.
Assuming an exponential distribution of $^{218}$Po$^+$ neutralization times with mean lifetime $\tau_\mathrm{n}$, the corresponding \RnPo charge branching fraction must vary with $\tau_\mathrm{n}$ as shown in Figure~\ref{fig:RnPoChargeBranchingForEachNeutralization} to produce the observed number of $^{218}$Po decays in the neutral search window.
%To produce the number of observed $^{218}$Po decays in the neutral search window, the \RnPo charge branching fraction must vary with the $^{218}$Po$^+$ neutralization lifetime $\tau_\mathrm{n}$ as shown in Figure~\ref{fig:RnPoChargeBranchingForEachNeutralization},
%assuming an exponential distribution for the charge neutralization time.

\begin{figure}[t]
    \centering
    \includegraphics[width=\columnwidth]{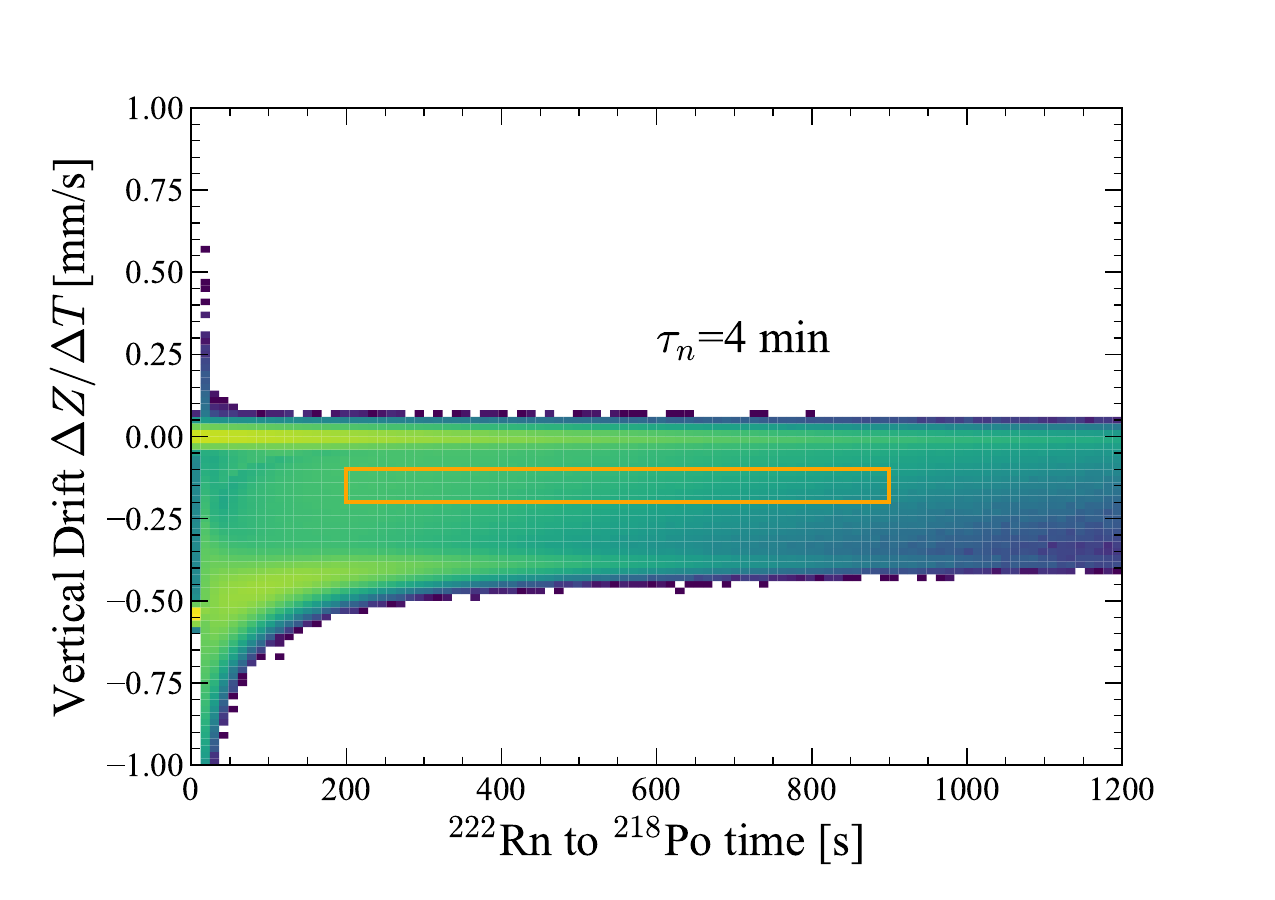}
    \includegraphics[width=\columnwidth]{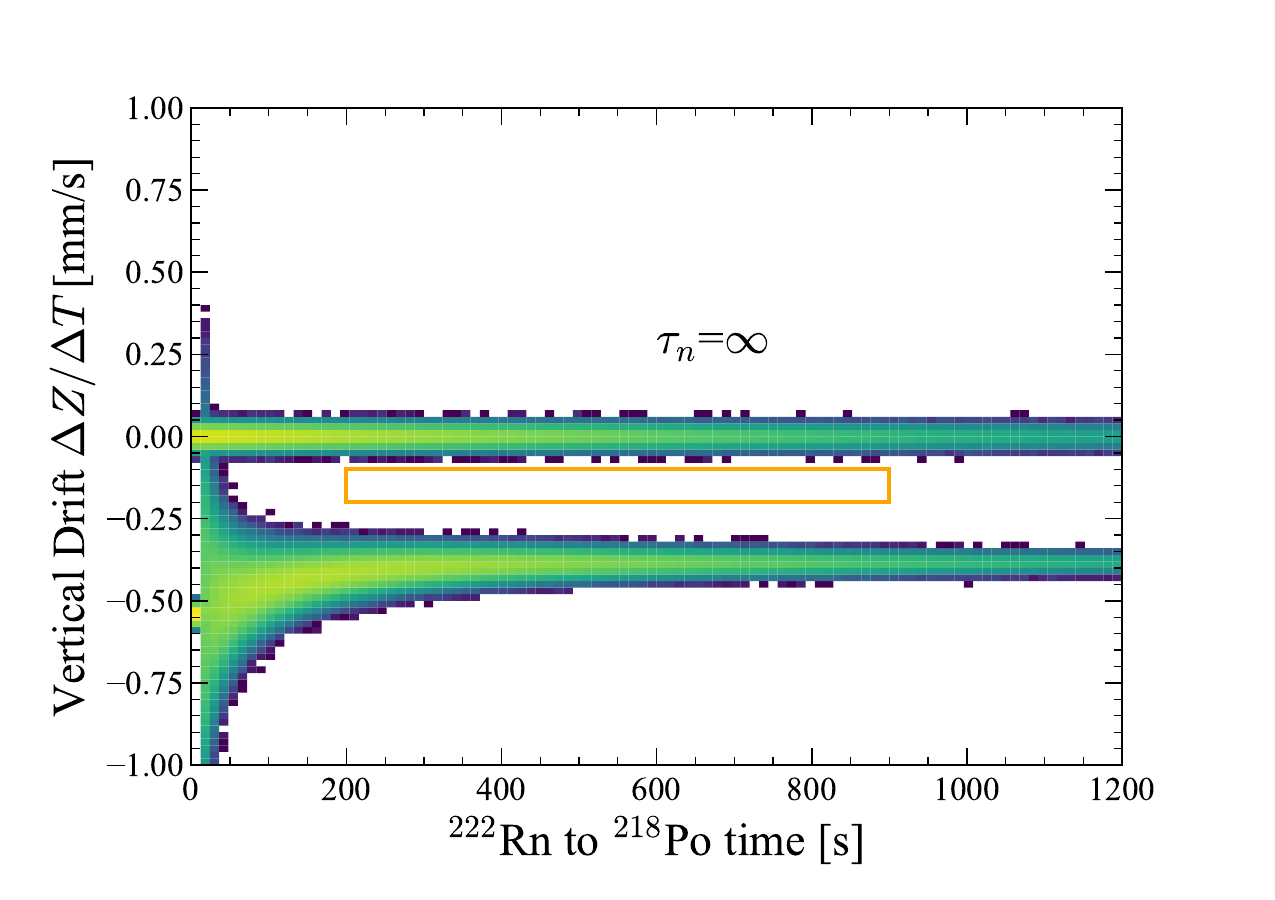}
    \caption{Two Monte-Carlo simulations of \RnPo pairing with a mean neutralization time of 4~minutes (above) and no neutralization (below). The simulated number of pairs in the orange box between the neutral and charged progeny bands (9924 pairs for $\tau_\mathrm{n}$=4~min, 0 pairs for $\tau_\mathrm{n}$=$\infty$) is compared to the 1065 excess pairs observed in data to constrain the charge mean neutralization time.} 
    \label{fig:SimNeutralizationExtremes}
\end{figure}

\begin{figure}[h]
    \centering
    \includegraphics[width=\columnwidth]{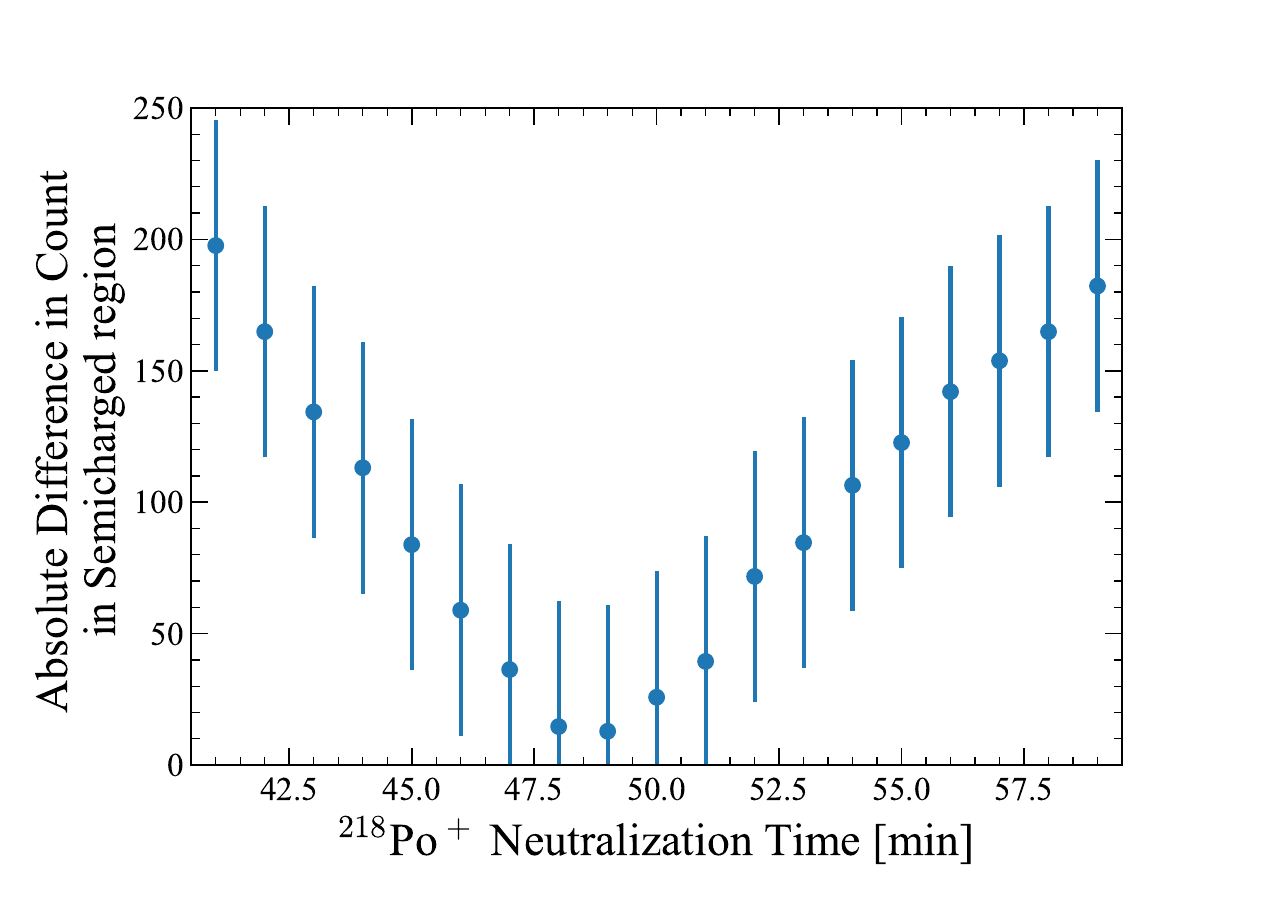}
    \caption{Absolute differences between the simulated number of pairs and the observed pair excess in the charge neutralization region indicated in Figure~\ref{fig:SimNeutralizationExtremes}, as a function of the simulated mean charge neutralization time. Error bars reflect the statistical uncertainty on the observed pair excess. }
    \label{fig:RnPoNeutralizaitonComparison}
\end{figure}

Given a hypothesized $\tau_\mathrm{n}$ and the corresponding inferred $^{222}$Rn--$^{218}$Po charge branching fraction, a simple charge-drift Monte-Carlo simulation can predict the distribution of pairs in $\Delta Z/\Delta T$ vs $\Delta T$ space.  Figure~\ref{fig:SimNeutralizationExtremes} shows the resulting distributions for the extreme cases of $\tau_\mathrm{n}=4$~min and $\tau_\mathrm{n}=\infty$.  Counting the pairs in the highlighted region (200~s~$<\Delta T<$~900~s and $-0.2$~mm/s~$<\Delta Z / \Delta T<$~$-0.1$~mm/s), where the pair rate is dominated by neutralized charged progeny, and comparing to the observed pair excess gives a best fit $\tau_\mathrm{n} = 49~\textrm{min}\pm4~\textrm{min}$ (see Figure~\ref{fig:RnPoNeutralizaitonComparison}), with a corresponding $^{222}$Rn--$^{218}$Po charge branching ratio of $0.49\pm0.01$, in agreement with the $0.50\pm0.03$ previously reported by EXO-200~\cite{EXO-200chargedions}.  The uncertainty on $\tau_\mathrm{n}$ is statistical, while the uncertainty on the charge branching fraction is dominated by progeny selection efficiency, which is in turn dominated by DAQ live time uncertainty.

The $^{218}$Po--$^{214}$Pb charge branching fraction can be similarly calculated from the observed \PoPb neutral-progeny pair excess, again requiring $\Delta T\in[200\,\textrm{s}, 900\,\textrm{s}]$ and assuming the same mean neutralization time for $^{214}$Pb$^+$ ions as was measured for $^{218}$Po$^+$ ions.  The $^{218}$Po--$^{214}$Pb charge branching of $0.48\pm0.12$ reported in Table~\ref{tab:ChargeBranching} is based on the $^{214}$Pb-dominated 18--75~keV$_{\textrm{ee}}$ region, with a large statistical uncertainty due to the low statistics in that region.  This is the first direct measurement of $^{218}$Po--$^{214}$Pb charge branching and supports the assumption in \cite{EXO-200chargedions} of equal charge branching for the $^{222}$Rn and $^{218}$Po alpha-decays.

The reported $^{214}$Pb--$^{214}$Bi charge branching fraction of $0.74\pm0.05$ is based on the \PoBiPo neutral-progeny pair excess with $\Delta T\in[200\,\textrm{s}, 900\,\textrm{s}]$. This result takes into account the number of semi-charged decay pairs appearing in the neutral search window as well as pairs with $^{214}$Pb$^+$ and/or $^{214}$Bi$^+$ neutralization (both assumed to have the same mean neutralization time as $^{218}$Po$^+$).  The charge states following the $^{218}$Po and $^{214}$Pb decays are assumed to be independent. The reported uncertainty on the $^{218}$Po--$^{214}$Pb charge branching fraction is not propagated; instead the $^{218}$Po--$^{214}$Pb charge branching fraction adopts the same value (and uncertainty) as the measured $^{222}$Rn--$^{218}$Po charge branching fraction.  The resulting $^{214}$Pb--$^{214}$Bi charge branching fraction is consistent with the $0.76\pm0.06$ reported by EXO-200~\cite{EXO-200chargedions}.  The higher charge branching fraction for the beta-decay is consistent with an interpretation where most ``neutral'' progeny are a product of prompt recombination at the event site, and the higher charge branching fraction of the beta-decay is a consequence of the lower recombination in electron tracks relative to alpha particle tracks. This interpretation also supports the assumption that the progeny charge state is independent of progenitor charge.

\end{document}